\documentclass[aps,twocolumn,amsmath,amssymb,groupedaddress,longbibliography,nofootinbib]{revtex4-2}
\usepackage{bbm}
\usepackage{mathrsfs}
\usepackage{amsmath}
\usepackage{amsfonts}
\usepackage[colorlinks=true,linkcolor=blue,citecolor=blue,urlcolor=blue,anchorcolor=blue]{hyperref}
\usepackage{graphicx,epstopdf}
\usepackage{subfigure}
\usepackage{epsfig}
\usepackage{dcolumn}
\usepackage{bm}
\usepackage{color}
\usepackage{natbib}
\usepackage{amssymb}
\usepackage{xcolor}
\usepackage{braket}
\usepackage{soul}
\usepackage{CJKutf8}

\makeatletter
\newenvironment{notocentry}{%
  \begingroup
  \let\addcontentsline\@gobblethree
}{%
  \endgroup
}
\makeatother

\begin{document}
\begin{CJK}{UTF8}{gbsn}

\title{Weak ergodicity breaking without nonthermal eigenstates}

\author{Boning Huang (黄泊宁)$^{1,2}$}
\author{Yongguan Ke (柯勇贯)$^{1,3}$}\email{keyg@szu.edu.cn}
\author{Li Zhang (张莉)$^{1}$}
\author{Ling Lin (林凌)$^{1}$}
\author{Chaohong Lee (李朝红)$^{1,3}$}\email{chleecn@szu.edu.cn}

\affiliation{$^{1}$Institute of Quantum Precision Measurement, State Key Laboratory of Radio Frequency Heterogeneous Integration, College of Physics and Optoelectronic Engineering, Shenzhen University, Shenzhen 518060, China}

\affiliation{$^{2}$Laboratory of Quantum Engineering and Quantum Metrology, School of Physics and Astronomy, Sun Yat-Sen University (Zhuhai Campus), Zhuhai 519082, China}

\affiliation{$^{3}$Quantum Science Center of Guangdong-Hong Kong-Macao Greater Bay Area (Guangdong), Shenzhen 518045, China}

\begin{notocentry}
\begin{abstract}
The typical mechanisms of ergodicity breaking in isolated interacting quantum systems, such as many-body localization and quantum many-body scars, originate from the nonthermal nature of the underlying eigenstates.
Here, in the absence of nonthermal eigenstates, we identify a mechanism for collective revivals of multiparticle Wannier states (MWSs) associated with nearly linear bands in a spatially modulated Bose-Hubbard lattice.
The MWSs, as superpositions of multiparticle Bloch states within individual energy bands, give rise to band-resolved Wannier-sector fragmentation.
The key idea is that spatially periodic modulation folds and separates energy bands of a simple lattice into several sub-bands, among which nearly linear sub-bands inherit the linear segments of the original bands.
Although multiparticle Bloch states satisfy the eigenstate thermalization hypothesis (ETH), the MWSs in the nearly linear band still exhibit long-lived collective revivals, due to emergent equally spaced energy levels.
Our work provides a route to weak ergodicity breaking in which long-lived revivals arise from spectral phase coherence among ETH-satisfying eigenstates rather than from scar-like nonthermal eigenstates.
%
\end{abstract}
\end{notocentry}
 
\begin{notocentry}
\maketitle
\end{notocentry}

\textit{Introduction.} 
Thermalization in isolated interacting quantum systems, erasing memory of generic initial states, deepens the understanding  of quantum statistical mechanics.
According to the eigenstate thermalization hypothesis (ETH)~\cite{PhysRevA.43.2046,PhysRevE.50.888,srednicki1999approach,rigol2008thermalization,Deutsch_2018}, this loss of memory originates from the fact that individual eigenstates already encode thermal behavior.
Long-lived memory retention therefore requires suppressing or bypassing thermalization, such as, integrable systems with an extensive number of conserved quantities~\cite{kinoshita2006quantum,PhysRevLett.98.050405,RevModPhys.83.863}, many-body localization in disordered systems~\cite{PhysRevLett.95.206603,BASKO20061126,ROS2015420,annurev:/content/journals/10.1146/annurev-conmatphys-031214-014726,annurev:/content/journals/10.1146/annurev-conmatphys-031214-014701,RevModPhys.91.021001}, and Stark many-body localization in tilted systems ~\cite{PhysRevLett.122.040606,doi:10.1073/pnas.1819316116,PhysRevB.102.054206,PhysRevA.103.023323,PhysRevLett.127.240502,lhsz-dkmq}.
In these mechanisms, nonthermal eigenstates extend over a broad energy range, and thus lead to global failures of ETH and strong ergodicity breaking that nonthermal dynamics occur for a broad class of initial states.

By contrast, while generic initial states thermalize, certain initial states retain long-lived memory in the weak ergodicity breaking.  
A prominent mechanism is quantum many-body scars ~\cite{bernien2017probing,turner2018weak,PhysRevB.98.155134,PhysRevLett.122.040603,PhysRevX.10.011055,PhysRevX.11.021021,doi:10.1126/science.abg2530,serbyn2021quantum,jepsen2022long,zhang2023many,PhysRevResearch.5.023010,annurev:/content/journals/10.1146/annurev-conmatphys-031620-101617,PhysRevLett.132.150401,pizzi2025genuine,HanPu2025044600}, in which periodic revivals originate from a small subset of nonthermal eigenstates embedded into ETH-satisfying eigenstates, while they become a vanishing fraction in the thermodynamic limit. 
Such nonthermal eigenstates, accompanied by anomalously low entanglement, generally originate from dynamical constraints, such as dipole conservation, strong tilted potentials, and density-dependent tunneling. 
Hilbert space fragmentation~\cite{PhysRevX.9.021003,hudomal2020quantum,PhysRevX.10.011047,PhysRevB.101.174204,doi:10.1142/9789811231711_0009,PhysRevX.12.011050,moudgalya2022quantum,PhysRevB.109.184313,Li,PhysRevX.15.011035,yang2025constructing,7j6x-74f1} provides another route to weak ergodicity breaking by decomposing the Hilbert space into dynamically disconnected sectors. 
Similarly, such fragmentation usually comes from dynamical constraints and supports scarred subspaces containing ETH-violating eigenstates.
In addition, weak ergodicity breaking can arise from isolated integrable sectors~\cite{PhysRevResearch.7.023099} or isolated groups of low-entanglement eigenstates~\cite{PhysRevB.106.035123}. 
In these scenarios, long-lived memory retention remains associated with ETH-violating nonthermal eigenstates.
%
%
This naturally raises a fundamental question: \textit{can weak ergodicity breaking arise without any ETH-violating eigenstates?}

\begin{figure}
    \centering
    \includegraphics[width=1\linewidth]{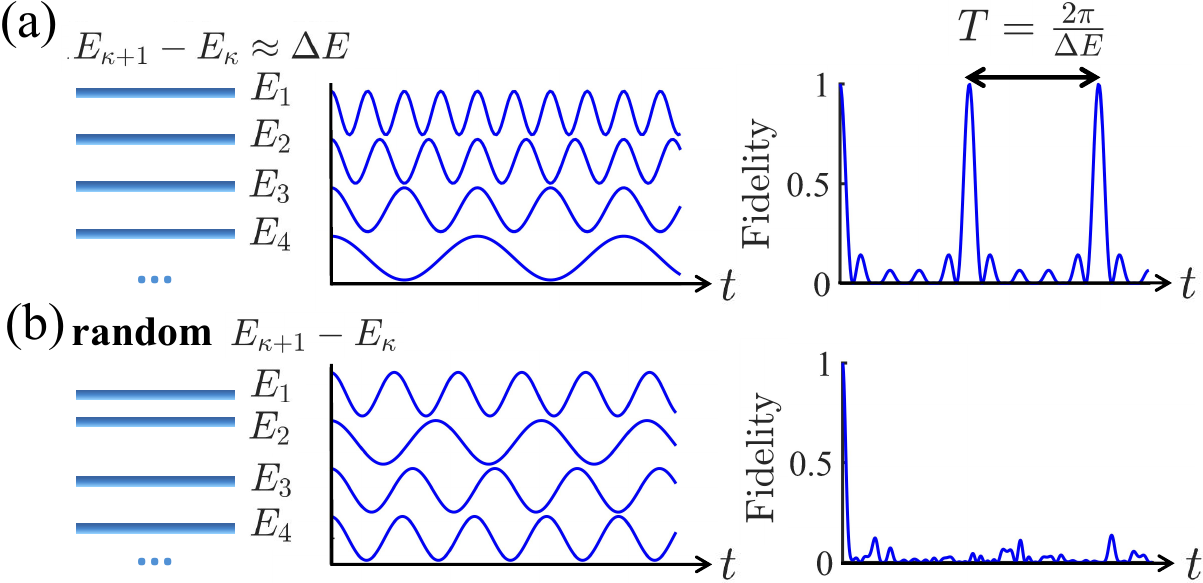}
    \caption{
    Schematics of (a) equally spaced energy levels in a linear band, which contribute to multiple frequencies and lead to periodic revival dynamics,
    and (b) random energy levels in a curved band, which contribute to random frequencies and lead to thermalization dynamics.
    }
    \label{fig:1}
\end{figure}

In this Letter, we show that weak ergodicity breaking can emerge from the periodically locked dynamical phases of ETH-satisfying eigenstates.
This process originates from the nearly equal energy spacings among these states, a phenomenon we term spectral phase coherence.
As a concrete realization, we consider a superlattice Bose-Hubbard model, in which the cotranslation symmetry allows multiparticle Bloch bands and multiparticle Wannier states (MWSs). 
In the multiparticle Wannier representation, the Hamiltonian naturally acquires a band-resolved block structure, in which each dynamical sector associates with a multiparticle energy band.
To produce nearly linear bands with nearly equal energy spacings, we apply spatially periodic modulation to fold multiparticle bands into sub-bands.
The MWSs in the nearly linear bands exhibit periodic revival dynamics [Fig.~\ref{fig:1}(a)], while those in curved bands with irregular energy spacings show dephasing [Fig.~\ref{fig:1}(b)].
The revivals are not due to nonthermal scar eigenstates, but originate from the phase locking in an ETH-satisfying spectrum.
Our work provides an unexplored mechanism for weak ergodicity breaking without the requirement of nonthermal eigenstates. 

\textit{Band-resolved Wannier-sector fragmentation.} 
Below we choose MWSs as basis and so that each multiparticle band forms an independent dynamical sector.
Under cotranslational symmetry, if all particles are simultaneously shifted by multiple unit cells, the system Hamiltonian remains invariant and the center-of-mass (c.m.) momentum $\kappa$ is a good quantum number.
The multiparticle Bloch states $|\psi_{m,\kappa}\rangle$ are the eigenstates with $\kappa$-dependent eigenenergies $E_{m,\kappa}$ forming multiparticle Bloch bands~\cite{PhysRevA.95.063630,PhysRevB.96.195134,qin2018topological,PhysRevA.101.023620,PhysRevResearch.5.013020,huang2024topological}.
In the basis of MWSs,
the Hamiltonian acquires a band-resolved block structure.
Without loss of generality, we consider $N$ particles in a lattice consisting of $L$ unit cells and assume $N$ and $L$ are coprime integers~\cite{Supplementary}.
A multiparticle Wannier state can be expressed as 
\begin{equation}
		|W_m(R)\rangle=\frac{1}{\sqrt{L}}\sum_{\kappa}e^{-i\kappa R}|\psi_{m,\kappa}\rangle,
\end{equation}
which is centered in the $R$th unit cell and uniformly occupies the $m$th multiparticle Bloch band.
The MWSs constitute a complete orthonormal basis 
$\langle W_m(R)|W_{n}(R')\rangle=\delta_{mn}\delta_{RR'}$, where $\delta$ is the Kronecker delta.
Therefore, the elements of Hamiltonian matrix are given by
\begin{equation}
    \label{eq4}
        \langle W_m(R)|\hat{H}|W_n(R')\rangle=\frac{1}{L}\sum_{\kappa}e^{i\kappa( R- R')} E_{m,\kappa} \delta_{mn},
\end{equation}
indicating that the Hamiltonian is block diagonal in the band index: Wannier states belonging to different multiparticle bands are dynamically decoupled~\cite{Supplementary}.
This structure is termed band-resolved Wannier-sector fragmentation. 
However, in a nonflat band sector, the off-diagonal matrix elements in the Wannier-center indices $R$ and $R'$ are generally nonzero, indicating that MWSs centered at different unit cells will couple with each other.
Unlike conventional constraint-induced Hilbert-space fragmentation, the sectors here arise from cotranslational symmetry rather than from dynamical constraints. 
This structure alone does not imply ergodicity breaking, because thermalization can still occur within each sector~\cite{PhysRevX.15.011035}.

If an initial state is prepared as a multiparticle Wannier state in the $m$th band, the evolved state $|W_m(R,t)\rangle=e^{-i\hat{H}t}|W_m(R,0)\rangle$ will always stay in the subspace of $m$th band, due to the fragmented Hilbert space~\cite{Supplementary}.
The expectation of an operator $\hat O$ at time $t$ can be given by
\begin{equation}
\begin{aligned}
      &\langle W_m(R,t)|\hat{O}|W_m(R,t)\rangle=
    \frac{1}{L}\sum_{\kappa}\langle \psi_{m,\kappa}|\hat{O}|\psi_{m,\kappa}\rangle\\&+\frac{1}{L}\sum_{\kappa \neq \kappa'}e^{i(E_{m,\kappa}-E_{m,\kappa'})t}e^{i(\kappa-\kappa')R}\langle \psi_{m,\kappa}|\hat{O}|\psi_{m,\kappa'}\rangle.
\end{aligned}
\end{equation}
Here, the first term is the diagonal contribution, which can be consistent with ETH.
The second term contains phase factors determined by all energy differences within the band.
In a generic curved band with irregular energy spacings, dephasing will cause the dynamics relax to the diagonal-ensemble value~\cite{rigol2008thermalization}. 
However, if the band has an approximately equal-spacing structure, the phases will be periodically locked and thus collective revivals appear.
%
This can be analyzed by the fidelity between the initial state and the instantaneous state
\begin{equation}
     |\langle W_m(R,0)|W_m(R,t)\rangle|^2
     =\frac{1}{L}+\frac{1}{L^2}\sum_{\kappa\neq\kappa'}e^{i(E_{m,\kappa}-E_{m,\kappa'})t}.
\end{equation}
%
%
%
%
%
Thus, long-lived revivals arise from spectral phase coherence within a Wannier sector and do not require nonthermal eigenstates.
In generic simple lattices, multiparticle bands $E_{m,\kappa}$ usually have irregular energy spacings, leading to dephasing and thermalization. 
A natural route to revivals is to engineer nearly linear multiparticle bands $E_{m,\kappa}\propto |\kappa|$ within a selected sector, thereby ensuring the MWSs periodically return to their initial states.
While such linear dispersion can be engineered in single-particle bands using long-range hopping, it does not generally survive in interacting multiparticle bands~\cite{Supplementary}.
Below we show that spatially periodic modulation offers a robust way to generate nearly linear multiparticle sub-bands in an interacting lattice.

\textit{Band folding in a superlattice Bose-Hubbard model.} 
We consider a superlattice Bose-Hubbard model with spatially periodic modulations,
\begin{equation}
        \hat{H}=-\sum_{j}J_j(\hat{a}_{j+1}^{\dag}\hat{a}_j+{\rm H.c.})+\sum_{j}V_j\hat{n}_j+\frac{1}{2}\sum_{j}U_j\hat{n}_j(\hat{n}_j-1).
\end{equation}
Here, $\hat{a}_{j}^{\dag}$ ($\hat{a}_j$) and $\hat{n}_j=\hat{a}_{j}^{\dag}\hat{a}_j$ are bosonic creation (annihilation) operators and particle number operators at the $j$th site, respectively.
We assume the system has $M$ lattice sites and $N=\sum_j \langle\hat{n}_j\rangle$ particles under the periodic boundary conditions.
We focus on considering interaction modulation $U_j=U_0+\delta_U g(j)$ and keep $J_j=J_0$ and $V_j=0$ in the main text; see Supplemental Material for the cases of hopping and onsite potential modulations.
For a $d$-period superlattice, we have $g(j)=g(j+d)$.
This model can be readily realized in various experimental platforms, such as ultracold atoms~\cite{walter2023quantization,ke2023topological} and superconducting circuits~\cite{ZiyuTao2025033202}.
In these platforms, the onsite interaction can be independently tuned by utilizing Feshbach resonance or controlling the anharmonicity of individual qubits, respectively.

\begin{figure}[t]
    \centering
    \includegraphics[width=0.98\linewidth]{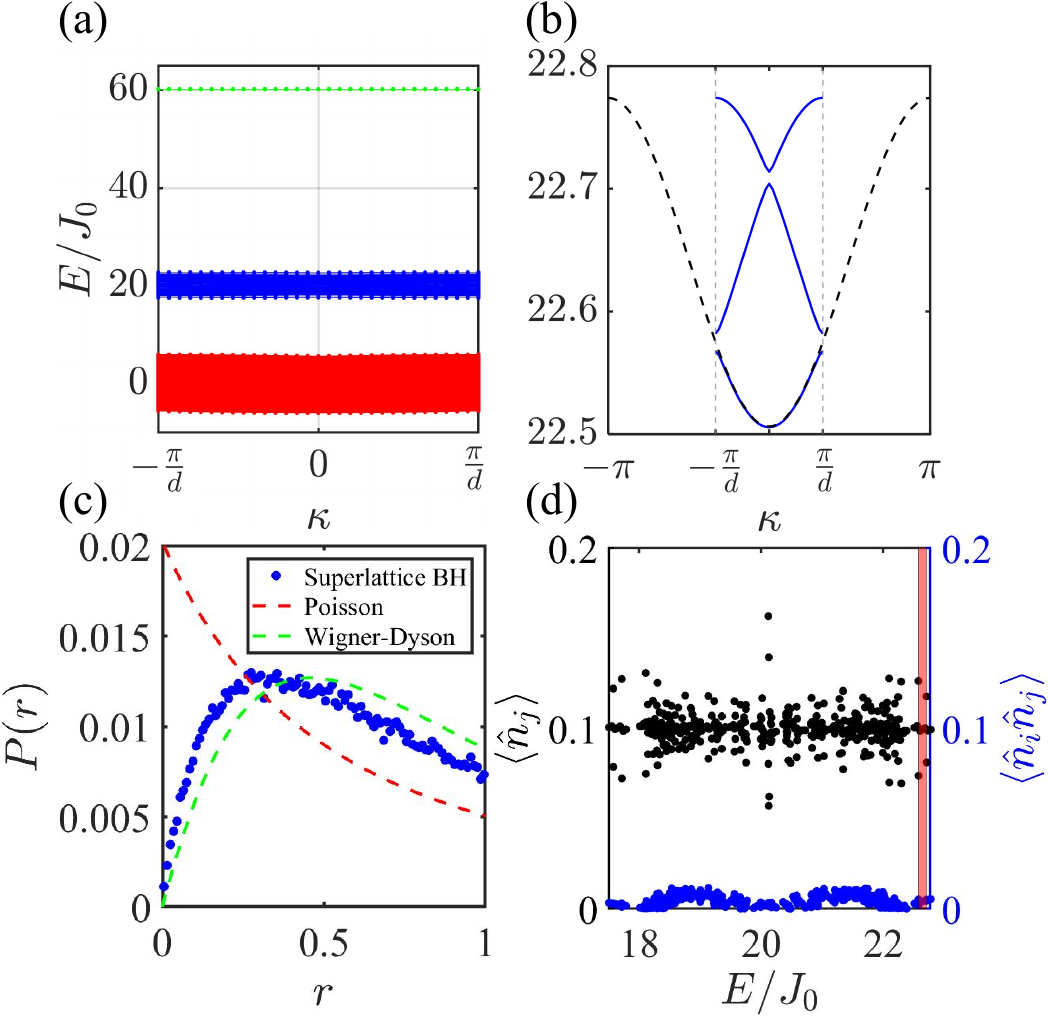}
    \caption{(a) Three-particle Bloch bands.
    (b) Appearance of nearly linear sub-bands (blue solid lines), folded from the highest dimer-monomer band in the original simple lattice (black dashed line).
    (c) Level statistics of the superlattice Bose-Hubbard model with weak onsite disorders (blue dots).
    Red dashed and green dashed lines denote Poisson distribution and Wigner-Dyson distribution, respectively.  
    (d) Expectation values of $\hat{n}_{j=15}$ (black) and $\hat{n}_{i=12}\hat{n}_{j=15}$ (blue) of dimer-monomer eigenstates as a function of energy. 
    Here $i$ and $j$ can be chosen as other sites, without changing physical picture.
    Values associated with a nearly linear band are highlighted by the red shaded region.
    Parameters are chosen as $J_0=1$, $U_0=20$, $d=3$, $g(j)\in\{-0.20,0.48,-0.30\}$.
    $\delta_U=0.05$ for (a) and blue lines in (b), $\delta_U=0$ for the black dashed line in (b), and $\delta_U=0.01$ for (c),(d).
    $M=90$ for (a),(b) and $M=30$ for (c),(d).
    }
    \label{fig:2}
\end{figure}

Under strong interactions ($|U_j| \gg |J_j|$), while particles at different sites can tunnel independently, particles at the same site will form bound states~\cite{winkler2006repulsively,valiente2008two,fukuhara2013microscopic}.
In Fig.~\ref{fig:2}(a), we show the three-particle energy bands with parameters $U_0=20$, $\delta_U=0.05$, $M=90$, $d=3$ and $g(j)\in\{-0.20, 0.48, -0.30\}$.
The spectrum separates into three manifolds and similar band-folding physics also appears in other choices of $g(j)$.
In addition to scattering states (red lines) and three-particle bound states (green lines), there exist dimer-monomer states (blue lines), which include exotic interaction-induced bound states in continuum~\cite{zhang2023stable,PhysRevLett.133.140202,PhysRevLett.133.193001}.

In a simple lattice system ($\delta_U=0$), the Brillouin zone ranges from $-\pi$ to $\pi$ and the highest dimer-monomer band (black dashed line) has a sinusoidal dispersion.
However, when a spatially period-three modulation is applied, the highest dimer-monomer band is folded three times into the reduced Brillouin zone $[-\pi/d, \pi/d)$.
Around the energy crossing points $\kappa=0,~\pm \pi/d$, gaps are opened by the applied modulation and their widths increase with the modulation strength, leading to three sub-bands. 
In Fig.~\ref{fig:2}(b), we show the highest three dimer-monomer sub-bands, in which the middle band is almost a linear band.
The linear middle sub-band comes from the linear part of the original highest dimer-monomer band.
Increasing the period $d$, there appear more linear bands with narrower band widths.
However, there is a fundamental trade-off between linear dispersion and band-gap width~\cite{Supplementary}. 
This is because: (i) weak modulation can maintain sharp band edges and linear dispersion, whereas (ii) strong modulation opens large gaps but also induces curvature of the sub-bands.
Therefore, a moderate modulation strength represents an optimal compromise, preserving linear dispersion while still opening a finite band gap.
In this way, a nearly linear segment of an otherwise irregular multiparticle band can be isolated.

We next verify that the eigenstates forming the nearly linear band show no ETH violation.
To rule out an integrability-based mechanism, we first examine level statistics.
It is well known that the integrable points of the bare Bose-Hubbard model are no hopping or no interaction~\cite{Kolovsky2004QuantumCI}.
Except these two limit cases, the Bose-Hubbard model is generally non-integrable.
We calculate the level statistics of the considered dimer-monomer states of the superlattice Bose-Hubbard model; see Fig.~\ref{fig:2}(c).
The diagnostic is implemented by the distribution $P(r)$ of the adjacent energy gap ratio~\cite{PhysRevB.75.155111}
\begin{equation}
    r_m=\frac{{\rm min}(\delta_m,\delta_{m+1})}{{\rm max}(\delta_m,\delta_{m+1})},
\end{equation}
where $\delta_m=E_{m+1}-E_m$ is the energy difference between nearest-neighboring energy levels.
A weak disordered term $\sum_jW_j\hat{n}_j$ is introduced to break possible symmetries of the system,
where $W_j/J_0$ are random numbers in the range from $-0.1$ to $0.1$, and $100$ sets of $W_j$ provide sufficient samples.
For comparison, we  also show the Poisson distribution $P_{\rm P}(r)=2/(1+r)^2$ (for integrable systems) and Wigner-Dyson distribution of Gaussian orthogonal ensembles $P_{\rm GOE}(r)=27(r+r^2)/4(1+r+r^2)^{5/2}$ (for chaotic systems)~\cite{PhysRevLett.110.084101}.
Owing to the strong interaction, the system approaches the no-hopping limit and thus does not perfectly conform to the Wigner-Dyson distribution. 
Nevertheless, the pronounced level repulsion at $r\rightarrow 0$ distinguishes it from integrable Poissonian systems~\cite{Kollath_2010}.

A more direct ETH diagnosis is provided in Fig.~\ref{fig:2}(d); see more details in Supplemental Material~\cite{Supplementary}.
We analyze the expectation values of local density $\hat{n}_{j}$ and density correlation $\hat{n}_i\hat{n}_j$ of dimer-monomer eigenstates for different energies, where the region of the nearly linear band is highlighted by the red shadow. 
The expectation values vary smoothly with the energy, with small fluctuations within a narrow energy window~\cite{rigol2008thermalization}.
The multiparticle Bloch states in the nearly linear band do not exhibit anomalous expectation values or unusually low entanglement~\cite{Supplementary}, which are  consistent with the ETH within our available finite-size diagnostics.
This means that, within our available finite-size diagnostics, the reviving Wannier sector is built from eigenstates not belonging to scar-like states.

\textit{Ergodicity breaking dynamics.} 
The nearly linear sub-band provides a coherent Wannier sector in which the multiparticle Bloch eigenstates have almost equally spaced energies.
A maximally localized MWS in this sector is  expected to exhibit collective oscillations.
To this end, we study time evolution of maximally localized MWSs in the nearly linear dimer-monomer band, $|\psi(t)\rangle=e^{-i\hat{H}t}|\psi(0)\rangle$ and calculate the density distribution $\langle \hat{n}_j(t)\rangle=\langle \psi(t)|\hat n_j |\psi(t)\rangle$ and fidelity $F(t)=|\langle \psi(0)|\psi(t)\rangle|^2$.
We also calculate the particle-partition entanglement entropy $S(t)=-\sum_\nu \lambda_\nu(t)^2 \textrm{ln} \lambda_\nu(t)^2$ and the corresponding Page value within the same Wannier sector~\cite{Page1993,poshakinskiy2021quantum,PhysRevLett.130.120401}.
Here, $\lambda_\nu(t)$ represents singular values given by singular value decomposition $|\psi(t)\rangle=\sum_{\nu}\lambda_{\nu}(t)|\varphi_{\nu}(t)\rangle|\chi_{\nu}(t)\rangle$ with $|\varphi_{\nu}(t)\rangle$ and $|\chi_{\nu}(t)\rangle$ respectively denoting single-particle and $(N-1)$-particle states~\cite{zhang2023stable,PhysRevLett.133.140202,Supplementary}. 
The Page value is the average entanglement entropy of random pure states in a given Hilbert space, and it serves as the expected value when the system becomes thermalized~\cite{Page1993}.
%
%
The associated suppression of information spreading is further characterized by out-of-time-ordered correlators (see details in Supplemental Material~\cite{Supplementary,larkin1969quasiclassical,kitaev2015simple,maldacena2016bound,swingle2018unscrambling,PRXQuantum.5.010201,cg3f-rggs}).

\begin{figure}
    \centering
    \includegraphics[width=1\linewidth]{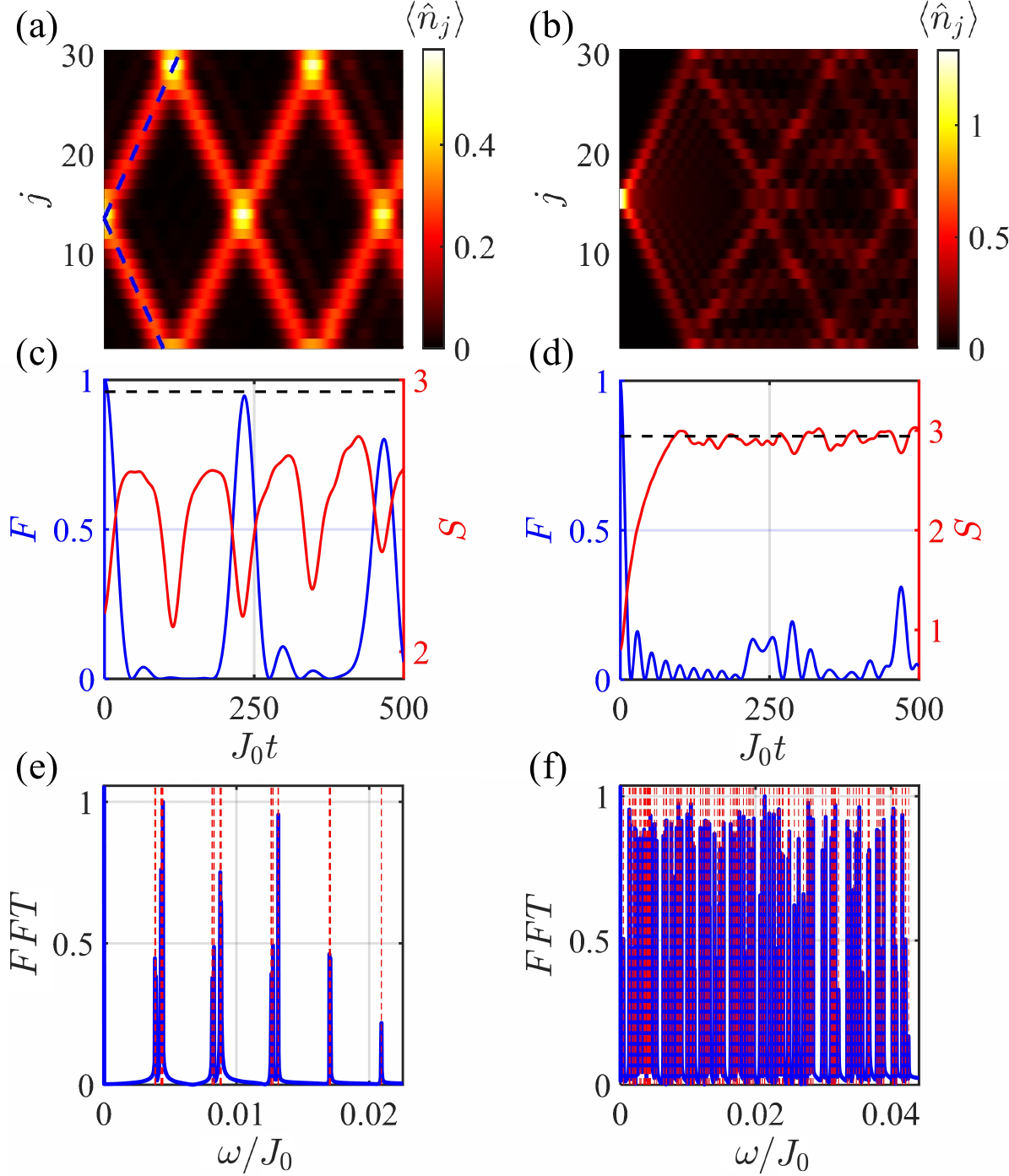}
    \caption{Dynamics of multiparticle Wannier states in different lattices: the superlattice (left panel) and the simple lattice (right panel). 
    (a, b) Time-evolution of density distribution.
    Blue lines in (a) denote the mean displacement predicted by the group velocities of the nearly linear band. 
    (c, d) Time-evolution of fidelity (blue) and entanglement entropy (red).
    Black dashed lines in (c) and (d) denote the Page value. 
    (e, f) Normalized fast Fourier transform (FFT) spectrum of fidelity in long-time dynamics.
    Vertical dashed red lines denote the frequencies given by the energy gaps in corresponding band.
    $\delta_U=0.01 (\delta_U=0)$ is chosen for superlattice (simple) lattice, while the other parameters are chosen as the same as those in Fig.~\ref{fig:2}(d).
    %
    }
    \label{fig:3}
\end{figure}

In Fig.~\ref{fig:3}, we show the time-evolution of maximally localized MWSs in the dimer-monomer bands.
In our calculations, the parameters are chosen the same as Fig.~\ref{fig:2}(d).
Under moderate spatial modulations, in Figs.~\ref{fig:3}(a) and (c), we show the dynamics of a maximally localized MWS in the nearly linear second highest dimer-monomer band. 
Since the subspace is perfectly decoupled from other parts of the whole Hilbert space, the numerical calculations can be performed within this subspace, thereby significantly reducing the computational cost~\cite{Supplementary}.
The wave packet splits into two components propagating with opposite group velocities $v=\partial E_{m,\kappa}/\partial \kappa$, and their recombination gives rise to fidelity revivals.
The entanglement entropy also oscillates periodically and remains far below the Page value.
The two components come back to their initial position when they meet each other twice.
Therefore the revival frequency for entanglement entropy are twice as the one for density distribution and fidelity.
This indicates that the dynamics does not explore the fully thermalized sector, despite being generated by ETH-satisfying eigenstates.
%
%
Nevertheless, the small fraction of non-equal energy spacings around band edges will lead to a small fraction of different group velocities. 
Although the slightly curved band edges cause transient damping of the oscillations, the maximally localized MWS exhibits long-term beat oscillations~\cite{Supplementary}. 
Generically, the collective revivals persist beyond the ideal MWS limit and remain robust against disorder: they appear in experimentally accessible Fock-state superpositions that significantly overlap the coherent Wannier sector, scale to systems with more particles, and manifest in both additional hybrid manifolds and coherent spectral structures (see Supplemental Material for details~\cite{Supplementary}).

\begin{figure}
    \centering
    \includegraphics[width=1\linewidth]{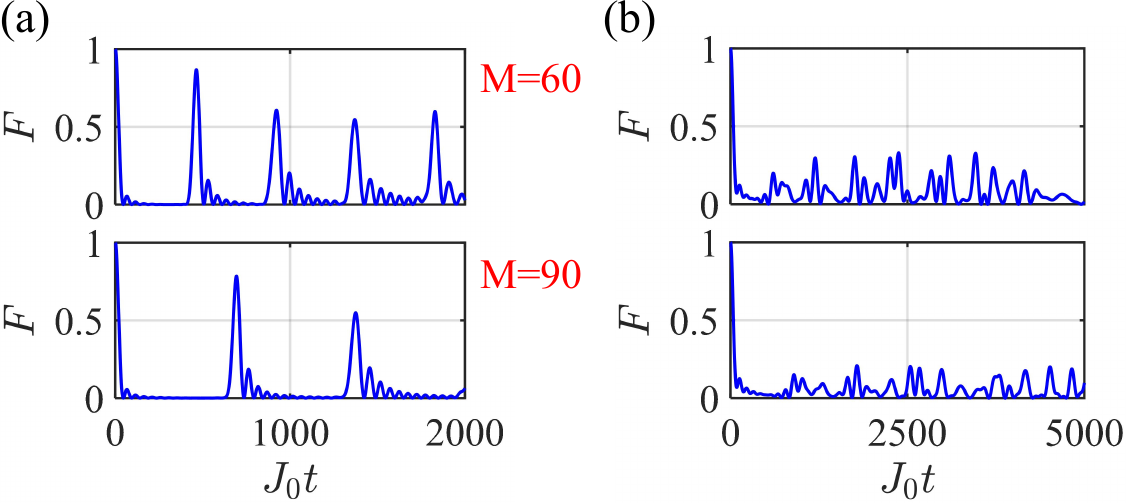}
    \caption{Time-evolution of fidelity in (a) the nearly linear dimer-monomer band and (b) curved highest dimer-monomer band for different system sizes: $M=60$ (top panel) and $M=90$ (bottom panel).
    Other parameters are the same as those in Figs.~\ref{fig:3}(a,~c). 
    }
    \label{fig:4}
\end{figure}

By contrast, in the simple lattice ($\delta_U=0$), the corresponding Wannier-sector dynamics loses coherence, because the band is strongly curved [Figs.~\ref{fig:3}(b),(d)] and the widely-distributed group velocities cause the wavepacket spread.
As evidenced by the fast Fourier transform (FFT) spectrum of fidelity for a long-time evolution,
there are peaks at finite commensurate frequencies for the superlattice nearly linear band [Fig.~\ref{fig:3}(e)] and many incommensurate frequencies for the simple lattice curved band [Fig.~\ref{fig:3}(f)].
As marked by the red dashed lines, the peaks in FFT spectrum match well with the energy gaps between multiparticle Bloch states. 
The revival (diffusive) dynamics originate from nearly equal (irregular) energy spacings in the associated bands.
The observed revivals are not simply finite-size recurrences,
since the coherent revivals in the nearly linear band remain visible for larger systems, with a longer period set by the decreasing level spacing [Fig.~\ref{fig:4}(a)].
Although periodic oscillations can also occur in small curved-band systems~\cite{Supplementary}, 
they are suppressed as the system size increases~[Fig.~\ref{fig:4}(b)].
%

\textit{Summary and Discussion.}
We have uncovered a route to weak ergodicity breaking that does not require nonthermal eigenstates.
Under moderate modulation, multiparticle Bloch bands can be folded into nearly linear sub-bands with approximately equal energy spacings.
Although constructed from ETH-satisfying multiparticle Bloch states, the MWSs in the nearly linear band exhibit long-lived periodic revivals.
These collective revivals originate from phase coherence among ETH-satisfying eigenstates, in  contrast to quantum many-body scars supported by nonthermal eigenstates.
This mechanism is expected to merge with thermalization in the thermodynamic limit,
because the spacing between adjacent energy levels vanishes, causing the revival period to diverge.
In this sense, the phenomenon also represents weak rather than strong ergodicity breaking.
Our scheme can be generalized to other translation-invariant quantum interacting systems, such as quantum spin chains and Fermi-Hubbard models.
In future, a promising extension is to generalize the spatial modulation to spatiotemporal modulation, which could pave the way for correlated space-time crystals with multiple periods. 
Our results identify spectral engineering of multiparticle energy bands as a route to coherent dynamics beyond the conventional scar paradigm.

The authors acknowledge useful discussions with Wenjie Liu, Dechi Peng, Jungeng Zhou, and Xinrui You. 
This work is supported by the Quantum Science and Technology - National Science and Technology Major Project (2025ZD0300800), the National Natural Science Foundation of China (92476201, 12275365, and 12175315), and the Guangdong Provincial Quantum Science Strategic Initiative (GDZX2305006, GDZX2405002 and GDZX2405003).
Li Zhang is supported by the National Natural Science Foundation of China (12305048) and Shenzhen Fundamental Research Project (JCYJ20230808105009018).

\providecommand{\noopsort}[1]{}\providecommand{\singleletter}[1]{#1}%
\begin{notocentry}

\end{notocentry}

\onecolumngrid
\clearpage

\begin{center}
	\noindent\textbf{\large{Supplementary material:}}
	\\\bigskip
	\noindent\textbf{\large{Weak ergodicity breaking without nonthermal eigenstates}}
	\\\bigskip
	\onecolumngrid
	
	Boning Huang (黄泊宁)$^{1,2}$, Yongguan Ke (柯勇贯)$^{1,3}$,$^*$ Li Zhang (张莉)$^{1}$, \\ Ling Lin (林凌)$^{1}$, and Chaohong Lee (李朝红)$^{1,3 \dag}$
	
	\small{$^1$\emph{Institute of Quantum Precision Measurement, State Key Laboratory of Radio Frequency Heterogeneous Integration, College of Physics and Optoelectronic Engineering, Shenzhen University, Shenzhen 518060, China}}\\
	\small{$^2$\emph{Laboratory of Quantum Engineering and Quantum Metrology, School of Physics and Astronomy, Sun Yat-Sen University (Zhuhai Campus), Zhuhai 519082, China and}}\\
	\small{$^3$\emph{Quantum Science Center of Guangdong-Hong Kong-Macao Greater Bay Area (Guangdong), Shenzhen 518045, China}}\\
\end{center}

\setcounter{section}{0}
\setcounter{subsection}{0}
\setcounter{figure}{0}
\setcounter{table}{0}
\setcounter{equation}{0}

\renewcommand{\thesection}{S\arabic{section}}
\renewcommand{\thesubsection}{\Alph{subsection}}
\renewcommand{\thefigure}{S\arabic{figure}}
\renewcommand{\thetable}{S\arabic{table}}
\renewcommand{\theequation}{S\arabic{equation}}
\renewcommand{\theHsection}{supp.\arabic{section}}
\renewcommand{\theHsubsection}{supp.\arabic{section}.\arabic{subsection}}
\renewcommand{\theHfigure}{supp.\arabic{figure}}
\renewcommand{\theHtable}{supp.\arabic{table}}
\renewcommand{\theHequation}{supp.\arabic{equation}}

\begin{notocentry}
\tableofcontents
\end{notocentry}

\newpage
\section{Band-resolved Wannier-sector structure}
\subsection{Band-resolved Wannier-sector fragmentation}
\begin{figure}[!h]
    \centering
    \includegraphics[width=0.95\linewidth]{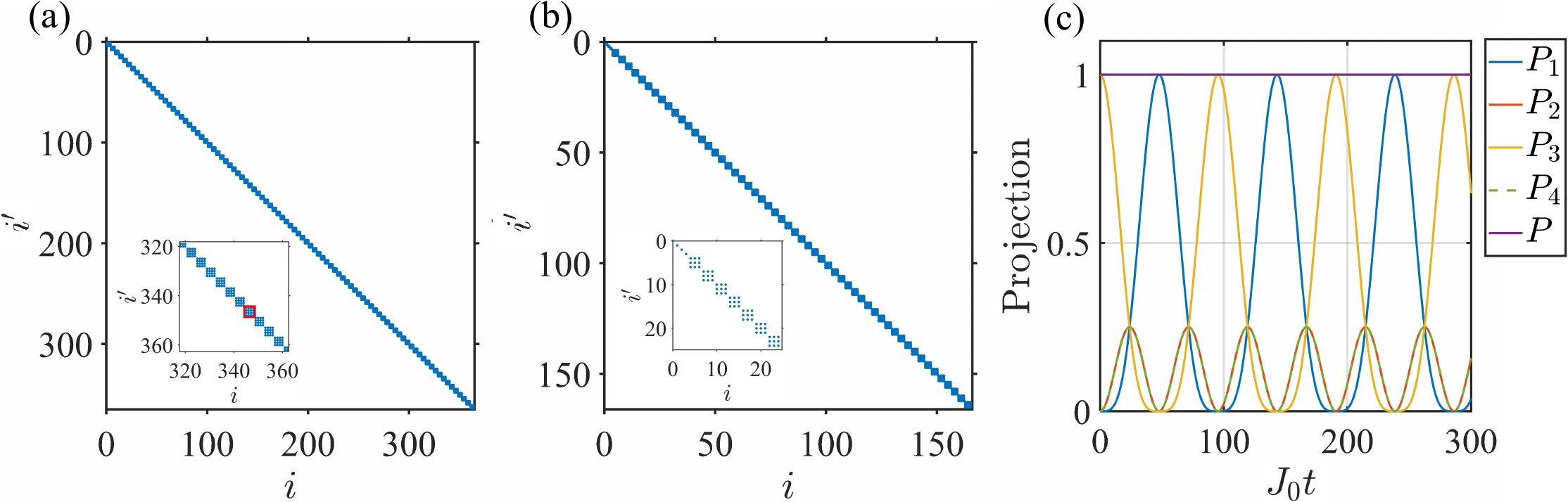}
    \caption{Sparse matrices of Hamiltonian spanned by basis of three-particle Wannier states with cell numbers (a) $L=4$ and (b) $L=3$.
    The basis are sorted by the band index. Insets show an enlargement of a part of the matrices.
    (c) Projection of evolved states onto three-particle Wannier states in corresponding Wannier sector as a function of time with cell number $L=4$.
    The initial state is a maximally localized MWS in the red box in (a).
    Other parameters are chosen as $J_0=1$, $\delta_U=0.01$, $U_0=20$, $d=3$, $g(j)\in\{-0.20, 0.48, -0.30\}$.}
    \label{fig:S1}
\end{figure}
In this section, we provide numerical evidence for the band-resolved Wannier-sector fragmentation induced by cotranslation symmetry.
In the multiparticle Wannier basis, matrix elements between states constructed from different Bloch bands vanish, so the Hamiltonian becomes block diagonal with respect to the band index.
When the number of particles $N$ and unit cells $L$ are coprime numbers, the Hilbert space is equally fragmented with each subspace sharing the same dimension.
In Fig.~\ref{fig:S1}(a), we show an example of the equally fragmented Hilbert space structure in the Wannier state basis when $N=3$ and $L=4$.
When $N$ and $L$ are not coprime numbers, the Hilbert space can be non-equally fragmented because of fewer energy levels in some bands.
The Wannier states in these bands are given by
\begin{equation}
		|W_n(R)\rangle=\frac{1}{\sqrt{L'}}\sum_{\kappa}e^{-i\kappa R}|\psi_{n}(\kappa)\rangle,
\end{equation}
where there are $L'$ ($L'<L$) values for $R$ and $\kappa$.
Nevertheless, they can still span a subspace decoupled with those of other bands; see sparse matrix of Hamiltonian in Fig.~\ref{fig:S1}(b) with $L=3$ and $N=3$. 
In the case of $L=6$, there will be $189$ $(6\times 6)$ blocks and three $(2\times 2)$ blocks.
Because of the band-resolved Wannier-sector structure, the dynamics of initial states in a given sector will be confined to such subspace.
In Fig.~\ref{fig:S1}(c), we present the projection onto corresponding sector as a function of time when evolving from a maximally localized multiparticle Wannier states (maximally localized MWS) in the  marked by red box in Fig.~\ref{fig:S1}(a),
which correspond to the second highest dimer-monomer band.
There are four Wannier states $|W_n(R)\rangle$ in the band, where the index $R$ ranges over $\{1, 2, 3, 4\}$.
During the dynamic process, 
projections of the evolved state onto these four maximally localized MWSs are defined as 
\begin{equation}
    P_R=|\langle W_n(R)|\psi(t)\rangle|^2.
\end{equation}
The state redistributes only among the four Wannier states in the selected sector, while the total projection onto this sector $P=\sum_RP_R$ remains unity. 
It directly verifies the dynamical isolation of the band-resolved Wannier sector.
Parameters are chosen as $J_0=1$, $\delta_U=0.01$, $U_0=20$, $d=3$, and $g(j)\in\{-0.20,0.48,-0.30\}$.

\subsection{Diagnostic of the band-resolved Wannier-sector fragmentation}
The degree of the band-resolved Wannier-sector fragmentation can be similarly measured by the ratio between the dimension of the largest fragment $D_{\rm frag}$ and total space $D_{\rm total}$~\cite{7j6x-74f1_SM}.
Here, $D_{\rm frag}$ is the number of unit cells $L$, because there are at most $L$ energy levels in a multiparticle Bloch band due to the cotranslation symmetry, and  
\begin{equation}
    D_{\rm total}=\begin{pmatrix}
    {N+M-1}\\N
\end{pmatrix}
\end{equation}
with particle number $N$ and system size $M$.
In the $d$-period superlattice, $L=M/d$.
After simplification, one can obtain 
\begin{equation}
    \frac{D_{\rm frag}}{D_{\rm total}}=\frac{N!M!}{d(N+M-1)!}.
\end{equation}
To analyze the thermodynamic scaling, assuming $N/M=s$ and considering $M, N\rightarrow \infty$，stirling approximation can be implemented that 
\begin{equation}
    \frac{N!M!}{(N+M-1)!}= \frac{\sqrt{2\pi MN(M+N)}M^MN^N}{(M+N)^{M+N}}.
\end{equation}
Then, we take the logarithm for analysis that
\begin{equation}
    {\rm ln}\frac{M^MN^N}{(M+N)^{M+N}}=M(s{\rm ln}s-(s+1){\rm ln}(s+1)).
\end{equation}
So, in the large $M, N$ limit, the ratio
\begin{equation}
     \frac{D_{\rm frag}}{D_{\rm total}} \sim \sqrt{M^3s(1+s)}e^{M(s{\rm ln}s-(s+1){\rm ln}(s+1))}.
\end{equation}
Due to $s{\rm ln}s-(s+1){\rm ln}(s+1)<0$, this ratio will become exponentially small in $M\rightarrow \infty$.
On the other hand, the number of fragments can be approximately estimated by 
\begin{equation}
    N_{\rm frag}=\frac{\begin{pmatrix}
    {N+M-1}\\N
\end{pmatrix}}{L},
\end{equation}
which is the reciprocal of $D_{\rm frag}/D_{\rm total}$.
It is because most Fock states return back to themselves after $L$ times cotranslation, leading to the dimension $L$ of the fragments.
Therefore each Wannier sector occupies an exponentially small fraction of the full Hilbert space in the thermodynamic limit. 
This scaling is consistent with the weak nature of the ergodicity breaking dynamics discussed in the main text.

\subsection{Projected Hamiltonian in Wannier sector}
\begin{figure}[!h]
    \centering   \includegraphics[width=0.8\linewidth]{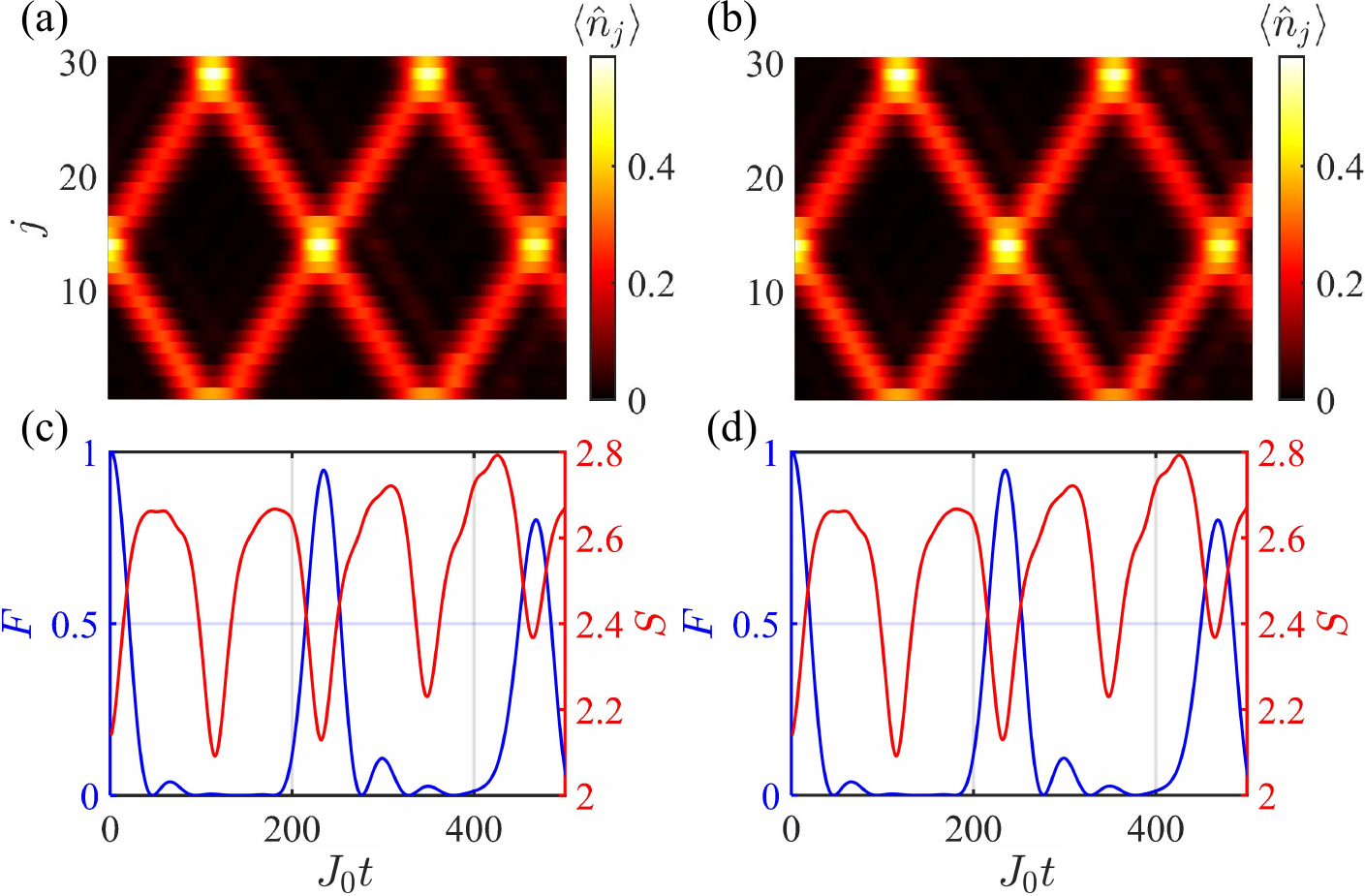}
    \caption{Comparison between the dynamics calculated in the total space (left panel) and the Wannier sector (right panel). 
    (a) and (b) The density distribution versus time. 
    (c) and (d) The fidelity (blue line) and entanglement entropy (red line) versus time.
    Parameters are $J_0=1$, $\delta_U=0.01$, $U_0=20$, $d=3$, $g(j)\in\{-0.20, 0.48, -0.30\}$, $M=30$.}
    \label{Effec}
\end{figure}
As the dynamics of multiparticle maximally localized MWSs are confined within the band-resolved Wannier sectors, and different Wannier sectors are exactly decoupled, 
the projected Hamiltonian gives the exact dynamics for an initial state inside a given sector, which can be written as
\begin{equation}
    \hat{H}_m=\sum_{R,R'}\langle W_m(R)|\hat{H}|W_m(R')\rangle|W_m(R)\rangle\langle W_m(R')|.
\end{equation}
Here, 
\begin{equation}
\langle W_m(R)|\hat{H}|W_m(R')\rangle=\frac{1}{L}\sum_{\kappa}e^{i\kappa(R-R')}E_{m,\kappa}
\end{equation}
is the coupling strength between multiparticle maximally localized MWSs centered at the $R$ and $R'$ cells.
Since there are only $L$ unit cells, the Hilbert space of the projected Hamiltonian is dramatically reduced.
Considering a maximally localized MWS in the second highest dimer-monomer band as the initial state,
the time evolution calculated in both the total space and the sector are presented in the left panel and right panel of Fig.~\ref{Effec}, respectively.
The two calculations are numerically indistinguishable.
Parameters are chosen as $J_0=1$, $\delta_U=0.01$, $U_0=20$, $d=3$, $g(j)\in\{-0.20, 0.48, -0.30\}$, $M=30$.
The agreement not only confirms the exact confinement of the dynamics within the selected Wannier sector,
but also provides a powerful numerical method to save computational resources.
For $N$ bosons, the full Hilbert space dimension 
$\begin{pmatrix}
    N+M-1\\N
\end{pmatrix}$
is reduced to at most $L$, 
which is greatly beneficial to numerical simulations of many-body systems.

\section{Engineering nearly linear multiparticle sub-bands}
\subsection{Absence of multiparticle linear band with long-range hopping}
\begin{figure}[!h]
    \centering
    \includegraphics[width=1\linewidth]{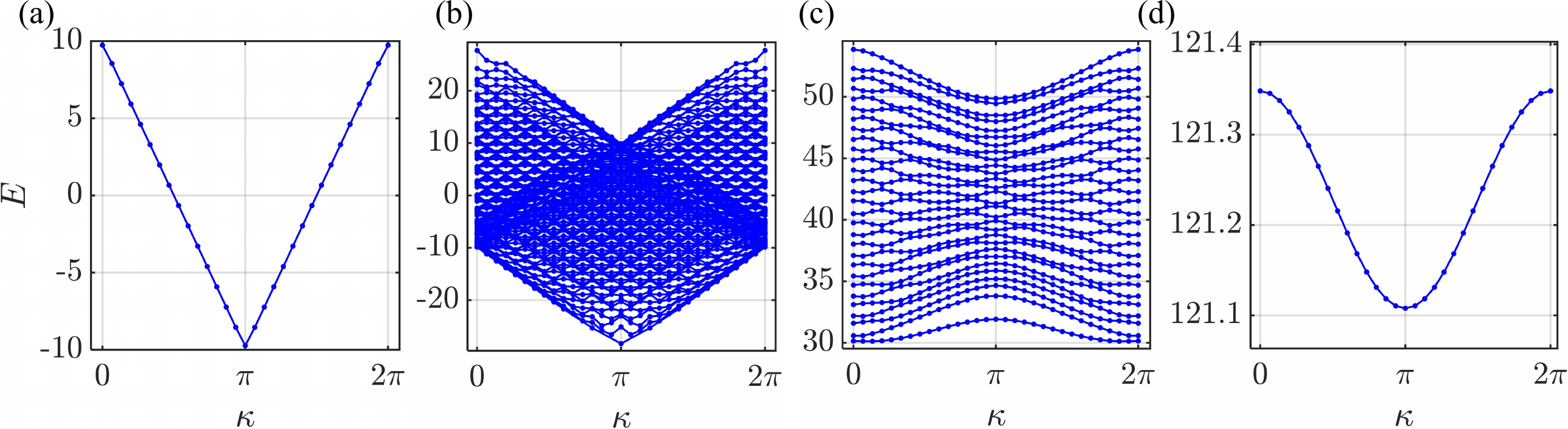}
    \caption{(a) Linear single-particle band, (b) the three-particle scattering-state bands,  (c) dimer-monomer bands, and (d) three-particle bound-state band under the long-range hopping. 
    The parameters are $J_n=((-1)^n-1)/n^2$, $U=40$, and $M=30$.
    }
    \label{Long_distance}
\end{figure}
Here we use long-range hopping as a counterexample to show that engineering a linear single-particle dispersion is not sufficient to obtain a linear multiparticle band.
Long-range hopping can be chosen to produce an exactly V-shaped single-particle dispersion.
We consider the Bose-Hubbard model with long-range hopping,
\begin{equation}
\hat{H}=-\sum_{j,n}J_n(\hat{a}_j^{\dagger}\hat{a}_{j+n}+{\rm H.c.})+\frac{1}{2}U\sum_j{\hat{n}_j(\hat{n}_j-1)},
\end{equation}
with $J_n$ to be determined
and the periodic boundary condition is adopted.
When considering single particle, the energy band is given by 
\begin{equation}
    E(k)=-2\sum_nJ_n{\rm cos}(kn).
\end{equation}
Because of the form of Fourier series, $E(k)$ can be designed as $E(k)=c|k|$ via engineering $J_n$.
Through inverse Fourier transformation, one can obtained 
\begin{equation}
    J_n\propto\frac{(-1)^n-1}{n^2}.
\end{equation}
Setting $J_n=((-1)^n-1)/n^2$, $U=40$, and the system size $M=30$,
Fig.~\ref{Long_distance}(a) shows the single-particle band, which is indeed linear.
Although the single-particle dispersion is linear, the interacting three-particle bands are strongly nonlinear, as shown in Figs.~\ref{Long_distance}(b),(c),(d), 
where there are scattering-state, dimer-monomer, three-particle bound-state bands, respectively.
Thus, the multiparticle equal-spacing structure must be engineered at the level of multiparticle band, rather than the single-particle dispersion.

\subsection{Linear band with modulation of hopping and onsite potential}
\begin{figure}[!h]
    \centering
    \includegraphics[width=0.8\linewidth]{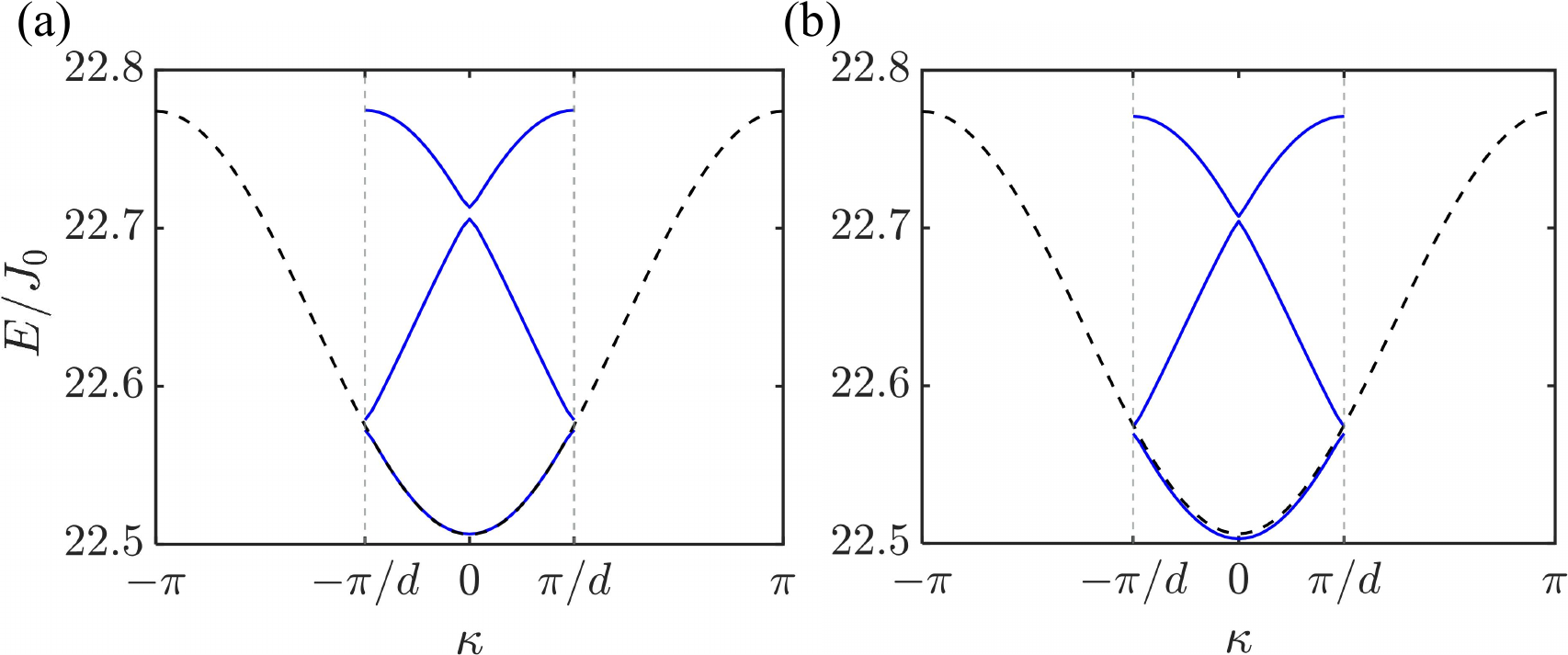}
    \caption{Sub-bands under the period modulation of (a) hopping and (b) onsite energy (blue solid lines) originated from the highest dimer-monomer band in the simple lattice (black dashed lines).
    Parameters are chosen as 
    $(J_0, V_0, U_0, \delta_J, \delta_V, \delta_U, d, M)=(1,0,20,0.01,0,0,3,90)$, $g(j)\in\{-0.83, 0.72, 0.16 \}$ for blue solid lines in (a), 
    $(J_0, V_0, U_0, \delta_J, \delta_V, \delta_U, d, M)=(1,0,20,0,0.003,0,3,90)$, $g(j)\in\{0.74, -0.88, -0.91\}$ for blue solid lines in (b),
    and $(J_0, V_0, U_0, \delta_J, \delta_V, \delta_U, d, M)=(1,0,20,0,0,0,3,90)$ for black dashed lines.
    }
    \label{othermodulation}
\end{figure}
The formation of nearly linear sub-bands is not specific to interaction modulation.
In this section, we show similar results under the modulation of hopping and onsite potential, respectively.
The spatial modulation with period $d$ will first fold the Brillouin zone from $(-\pi,\pi]$ in a simple lattice into $(-\pi/d,\pi/d]$.
Around the energy crossing points at $\kappa=0,~\pm \pi/d$, the modulation term provides off-diagonal elements in subspace spanned by $|\psi_m(\kappa)\rangle$ and $|\psi_{m+1}(\kappa)\rangle$, and the energy levels repel with each other, leading to energy avoided crossing and energy gap.
Nearly linear sub-bands can emerge when the modulation strength is moderate, regardless of the form of modulation.  
Fig.~\ref{othermodulation}(a) shows the bands in superlattice (blue lines) with $\delta_J=0.01$, $\delta_V=0$,  $g(j)\in\{-0.83, 0.72, 0.16\}$, 
which is folded from the highest dimer-monomer band in the simple lattice (black dashed line) with $\delta_J=0$, $\delta_V=0$.
Fig.~\ref{othermodulation}(b) shows them with $\delta_J=0$, $\delta_V=0.003$, $g(j)\in\{ 0.74, -0.88, -0.91
\}$.
Other parameters are chosen as $J_0=1$, $V_0=0$, $U_0=20$, $\delta_U=0$, $d=3$, $M=90$.
These results show that the modulation of hopping energy and onsite potential can generate nearly linear multiparticle sub-bands through the same band-folding mechanism.

\subsection{Breakdown of linear sub-bands under strong modulation}
\begin{figure}[!h]
    \centering
    \includegraphics[width=0.9\linewidth]{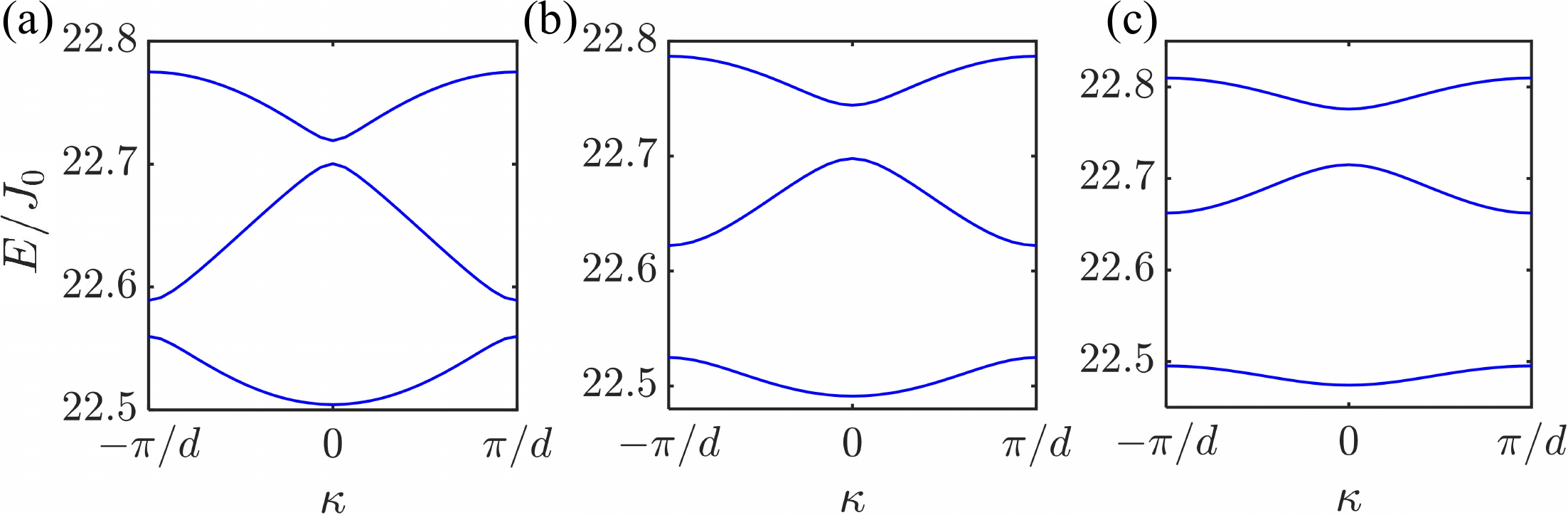}
    \caption{The highest three dimer-monomer bands under the period modulation of interaction (blue solid lines) with (a) $\delta_U=0.1$, (b) $\delta_U=0.3$ and (c) $\delta_U=0.5$.
    Other parameters are chosen as $J_0=1$, $U_0=20$, $d=3$, $g(j)\in\{-0.20, 0.48, -0.30\}$, and $M=90$}.
    \label{large_modu}
\end{figure}
There is a trade-off between opening a resolvable gap and preserving the near-linearity of the folded band.
We have to emphasize that the nearly linear bands can exist up to moderate modulation strength.
As spatial modulation strength further increases to strong regime, 
the nearly linear band will be gradually broken and turn to be curved band.
Figs.~\ref{large_modu}(a),(b),(c) show the highest three dimer-monomer bands under the modulation of interaction varying from $\delta_U=0.1$, $\delta_U=0.3$, and $\delta_U=0.5$, respectively.
Other parameters are set as $J_0=1$, $U_0=20$, $d=3$, $g(j)\in\{-0.20, 0.48, -0.30\}$, $M=90$.
With $\delta_U=0.1$, the middle band is still dominantly linear.
However, with the increased modulation strength,
the energy gap between the sub-bands become larger, 
the band edges become increasingly parabolic, and the middle band gradually loses its linearity. 
Therefore, the coherent equal-spacing structure required for revivals is optimized at intermediate modulation strengths.

\section{Thermalizing background and eigenstate diagnostics}
\subsection{Thermalization dynamics of dimer-monomer Fock states}
\label{Fockdynamics}
\begin{figure}[!h]
    \centering
    \includegraphics[width=0.98\linewidth]{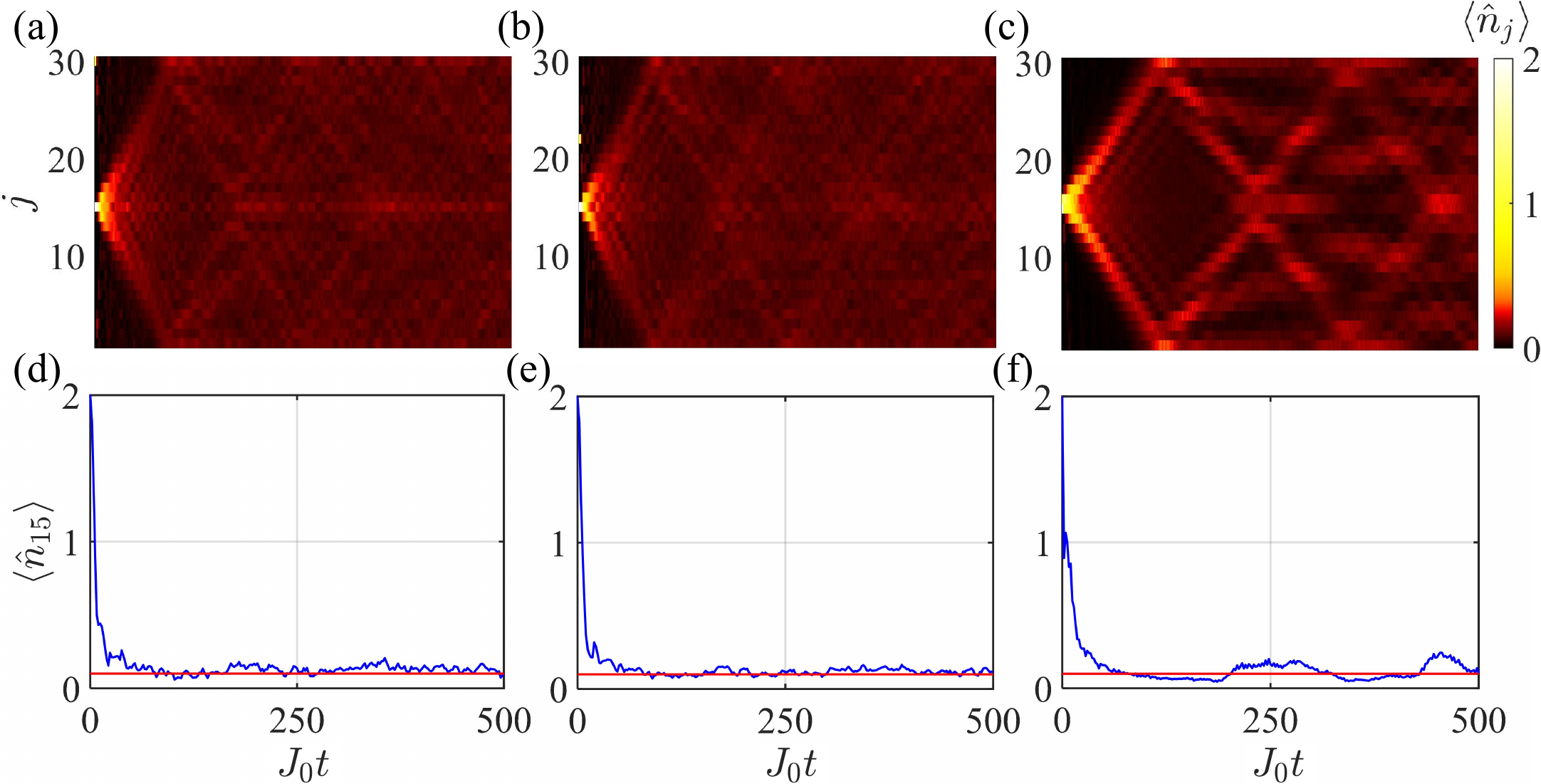}
    \caption{Thermalization dynamics of dimer-monomer Fock states.
    Density distribution as a function of time for initial states (a) $|2\rangle_{15}|1\rangle_{30}$, (b) $|2\rangle_{15}|1\rangle_{22}$, and
    (c) $|2\rangle_{15}|1\rangle_{16}$.
    Density at the $15$th site as a function of time for initial states (a) $|2\rangle_{15}|1\rangle_{30}$, (b) $|2\rangle_{15}|1\rangle_{22}$, and
    (c) $|2\rangle_{15}|1\rangle_{16}$.
    The red line indicates the predicted value of microcanonical ensemble.
    Parameters are chosen as $J_0=1$, $\delta_U=0.01$, $U_0=20$, $d=3$, $g(j)\in\{-0.20, 0.48, -0.30\}$, and $M=30$. 
    }
    \label{Fock}
\end{figure}
The revival dynamics discussed in the main text occurs within a background of otherwise thermalizing states.
In the main text, we have already shown the periodic revivals of multiparticle Wannier states in the nearly linear bands.
In this section, we consider the thermalization dynamics of the dimer-monomer Fock states.
These Fock states have energies in the same region as the Wannier states that exhibit revivals, but they are not prepared as coherent superpositions within the nearly linear sub-band.
Without loss of generality, we consider three dimer-monomer Fock states $|2\rangle_{15}|1\rangle_{30}$, $|2\rangle_{15}|1\rangle_{22}$, and $|2\rangle_{15}|1\rangle_{16}$ as initial states, 
where $|2\rangle_{j}|1\rangle_{j'}$ indicates that two particles are located at the $j$th site, and one particle is located at the $j'$-th site.
Under the periodic boundary condition, 
the relative distances between the two bound particles and the one independent particle are large, moderate, and small, respectively.  
Fig.~\ref{Fock} shows the thermalization dynamics for the  three states.
The initially localized density pattern rapidly spreads and relaxes toward an almost uniform distribution; see Figs.~\ref{Fock}(a),(b),(c).

To quantitatively test thermalization, 
we test whether the long-time average of the observables can be predicted by the microscopic ensemble average.
The predicted value of an observable corresponding to operator $\hat{O}$ is given by 
\begin{equation} \label{ensemble}
    O_{\rm mc}=(1/\Gamma_{E_0,\Delta E})\sum_{E_0-\Delta E<E_{m,\kappa}<E_0+\Delta E} \langle \psi_{m,\kappa}|\hat{O}|\psi_{m,\kappa}\rangle.
\end{equation}
$E_0=\langle \psi(0)|\hat{H}|\psi(0)\rangle$ is the energy of the initial state, 
$\Delta E$ is a small energy window constant which we choose as $\Delta E=1$,  
and $\Gamma_{E_0,\Delta E}$ is the number of eigenstates within this energy window.
Here, we calculate the observable $\langle\hat{n}_{15}\rangle=\langle\psi(t)|\hat{n}_{15}|\psi(t)\rangle$, which exhibits a rapid relaxation to the predicted value of the microcanonical ensemble; see Figs.~\ref{Fock}(d),(e),(f).
Parameters are chosen as $J_0=1$, $\delta_U=0.01$, $U_0=20$, $d=3$, $g(j)\in\{-0.20, 0.48, -0.30\}$, and $M=30$.
The long-time evolution of $\langle\hat{n}_{15}\rangle$ is well captured by the red lines which are calculated by Eq.~\eqref{ensemble}.
This confirms that generic dimer-monomer Fock states in the relevant energy window thermalize for local observables.
Hence, the revivals of maximally localized MWSs are not due to a nonthermal dimer-monomer energy window, but to the coherent equal-spacing structure of the selected nearly linear sub-band.

\subsection{Singular value decomposition of the $N$-particle state}
We use singular value decomposition to define the single-particle versus $(N-1)$-particle entanglement entropy used in the main text.
A $N$-particle state $|\psi \rangle$ expanded by the basis of Fock states can be reshaped to $|\psi \rangle=\sum_{j_1,j_2,...,j_N}\psi_{j_1,j_2,...,j_N}|\psi_{j_1,j_2,...,j_N}\rangle$, with the index $j_1,j_2,...,j_N\in [1,M]$ for the $N$ particles. 
Due to the symmetry of the bosonic particles, any swap of $j_1,j_2,...,j_N$ does not change the amplitude $\psi_{j_1,j_2,...,j_N}$.
So, we have a tensor whose elements are $\psi_{j_1,j_2,...,j_N}$, which can be reshaped from a $N$ dimensional tensor to a $M\times M^{N-1}$ matrix.
After that, we denote the elements of the matrix as $\tilde{\psi}_{j_1,r}$.
Performing singular value decomposition, $\tilde{\psi}_{j_1,r}$ can be represented by 
\begin{equation}
    \tilde{\psi}_{j_1,r}=\sum_{\nu}\lambda_{\nu} S_{j_1,\nu}W_{\nu, r}.
\end{equation}
$\lambda_{\nu}$ is the singular value, the column vector of $S_{j_1,\nu}$ represents a single-particle state, and the row vector of $W_{\nu, r}$ represent $(N-1)$-particle state. 
Then, the entanglement entropy can be given by 
\begin{equation}
    S=-\sum_{\nu}\lambda_{\nu}^2{\rm ln}\lambda_{\nu}^2,
\end{equation}
with the sigular values satisfy $\sum_{\nu}\lambda_{\nu}^2=1$.

\subsection{ETH diagnostics of $(N-1)$-bound-monomer eigenstates}
\begin{figure}[!h]
    \centering
        \includegraphics[width=0.8\linewidth]{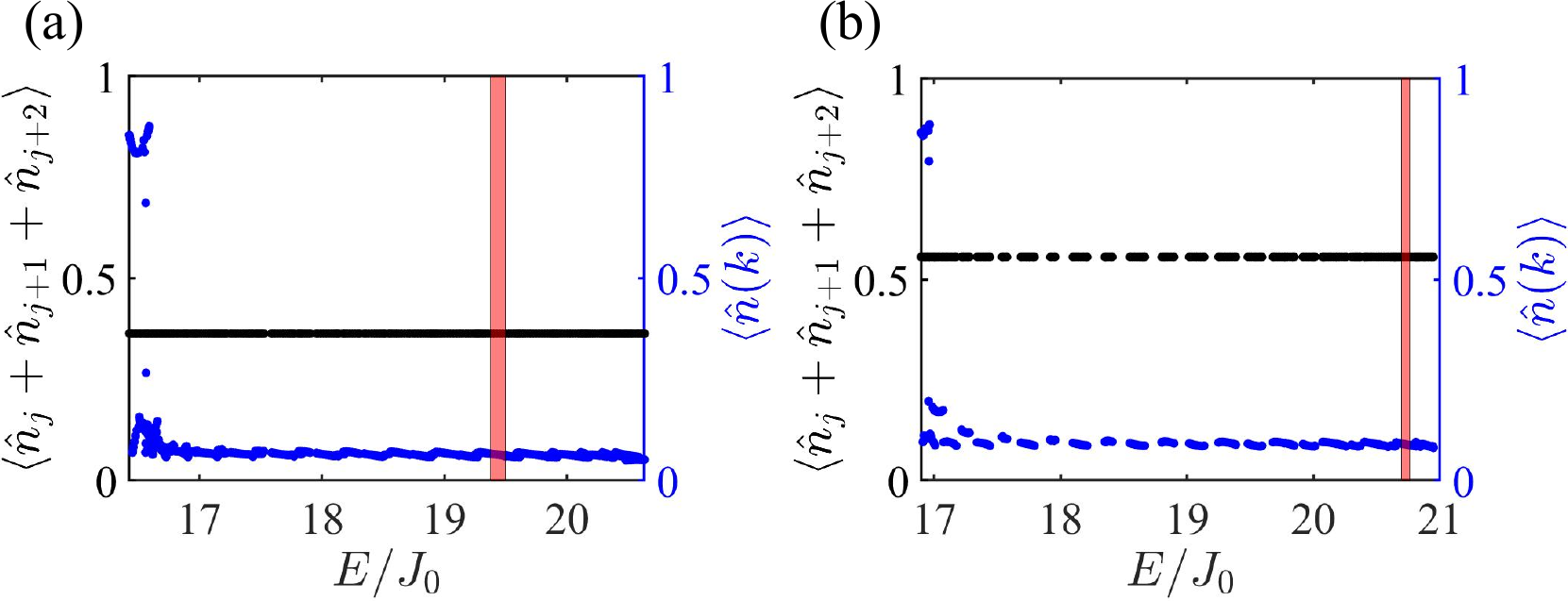}
    \caption{
    Eigenstates expectation values of local density operators (black) and single-particle momentum operator (blue) in (a) four-particle system and (b) five-particle system.
    Red shadows mark the values in selected nearly linear bands.
    Parameters are $(J_0, V_0, U_0, \delta_J, \delta_V, \delta_U, d, M)=(1,0,6,0.01,0,0,3,33)$, $g(j)\in\{ 0.39,-0.37,0.90\}$ for (a),
    and $(J_0, V_0, U_0, \delta_J, \delta_V, \delta_U, d, M)=(1,0,3,0.01,0,0,3,27)$, $g(j)\in\{ -0.75, 0.63, 0.81\}$ for (b).
    }
    \label{nk}
\end{figure}
We have presented an ETH diagnostic for the dimer-monomer eigenstates in the main text.
The local-density observable is a useful first diagnostic for detecting possible ETH-violating outliers. 
Nevertheless, because cotranslation symmetry constrains the density pattern of eigenstates [see black dots in Fig.~\ref{nk}], we also examine observables that are less directly fixed by this symmetry.
We show the eigenstates expectation values of single-particle momentum operator $\hat{n}(k)=(1/M)\sum_{i,j}e^{-ik(i-j)}\hat{b}_i^{\dagger}\hat{b}_j$; see blue dots in Fig.~\ref{nk}.
Here, $k=0$ is considered as an example.
In the $N$-particle system, there can be $(N-1)$-bound-monomer states with $N-1$ bound particles and an independent particle with enough interaction strength.
we also consider the $(N-1)$-bound-monomer eigenstates with $N=4$ and $N=5$ particles.  
The corresponding values in selected nearly linear bands is marked by red shadow, and the corresponding dynamics are shown in Section~\ref{more_particle}.
While there are individual large values, the values in the selected nearly linear bands do not form isolated anomalous outliers compared with nearby $(N-1)$-bound-monomer eigenstates.
The parameters are set as $(J_0, V_0, U_0, \delta_J, \delta_V, \delta_U, d, M)=(1,0,6,0.01,0,0,3,33)$, $g(j)\in\{ 0.39,-0.37,0.90\}$ for the four-particle case,
and $(J_0, V_0, U_0, \delta_J, \delta_V, \delta_U, d, M)=(1,0,3,0.01,0,0,3,27)$, $g(j)\in\{ -0.75, 0.63, 0.81\}$ for five particles.
Therefore, the revival dynamics shown in Section~\ref{more_particle} is not associated with anomalous eigenstate expectation values, but with the coherent energy-level structure of the selected band.

\subsection{Entanglement diagnostic of $(N-1)$-bound-monomer eigenstates}
\begin{figure}[!h]
    \centering
    \includegraphics[width=0.95\linewidth]{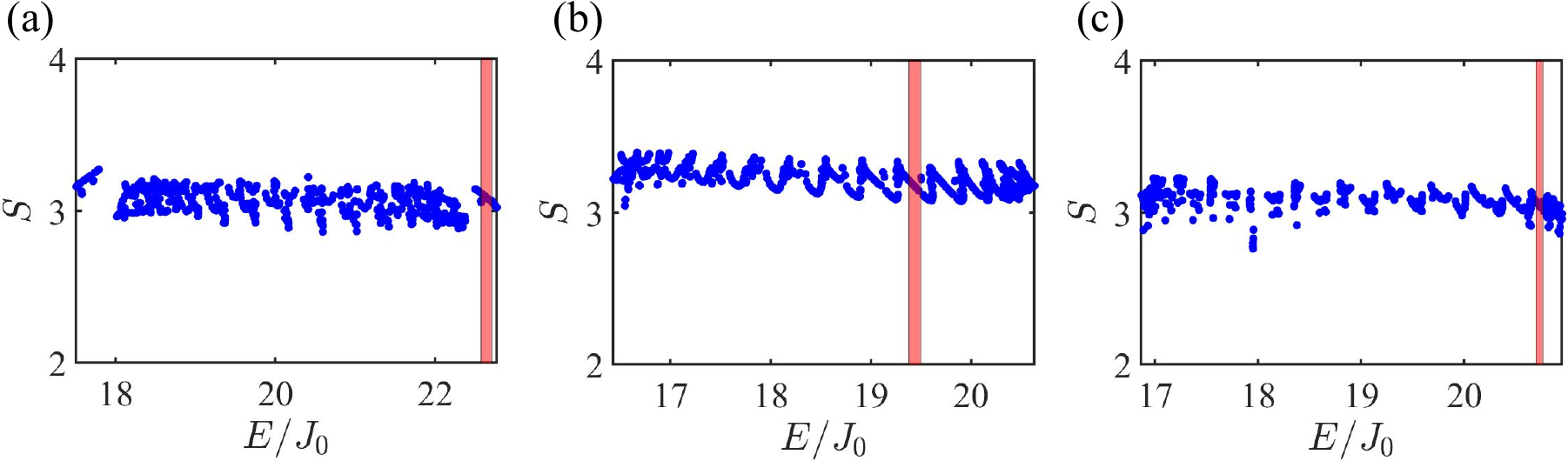}
    \caption{Entanglement entropy of $(N-1)$-bound-monomer eigenstates as a function of energy for (a) three-particle, (b) four-particle, and (c) five-particle systems.
    Red shadows mark the values in selected nearly linear bands.
    Parameters are $(J_0, V_0, U_0, \delta_J, \delta_V, \delta_U, d, M)=(1,0,20,0,0,0.01,3,30)$, $g(j)\in\{-0.20, 0.48,-0.30\}$ for (a),
    $(J_0, V_0, U_0, \delta_J, \delta_V, \delta_U, d, M)=(1,0,6,0.01,0,0,3,33)$, $g(j)\in\{ 0.39,-0.37,0.90\}$ for (b), 
    and $(J_0, V_0, U_0, \delta_J, \delta_V, \delta_U, d, M)=(1,0,3,0.01,0,0,3,27)$, $g(j)\in\{ -0.75, 0.63, 0.81\}$ for (c). 
    } 
    \label{entangle}
\end{figure}
In this section, we show the particle-partition entanglement entropy of $(N-1)$-bound-monomer eigenstates as a function of energy with different particle numbers in Fig.~\ref{entangle}.
The red shadows mark the values in selected nearly linear bands, and the corresponding dynamics are shown in Fig.~3 of the main text and Section~\ref{more_particle}.
Quantum many-body scar eigenstates commonly appear as anomalous low-entanglement eigenstates embedded in a thermal spectrum.
Here, we do not observe isolated low-entanglement outliers in the selected nearly linear bands.
The eigenstates forming the revival bands have entanglement entropies comparable to nearby dimer-monomer eigenstates. 
Thus, the revivals cannot be attributed to a scar-like set of anomalously low-entanglement eigenstates.

\section{Periodic revival dynamics and stability}
\subsection{Limited information spreading in the revival dynamics}
\begin{figure}[!h]
    \centering
    \includegraphics[width=0.7\linewidth]{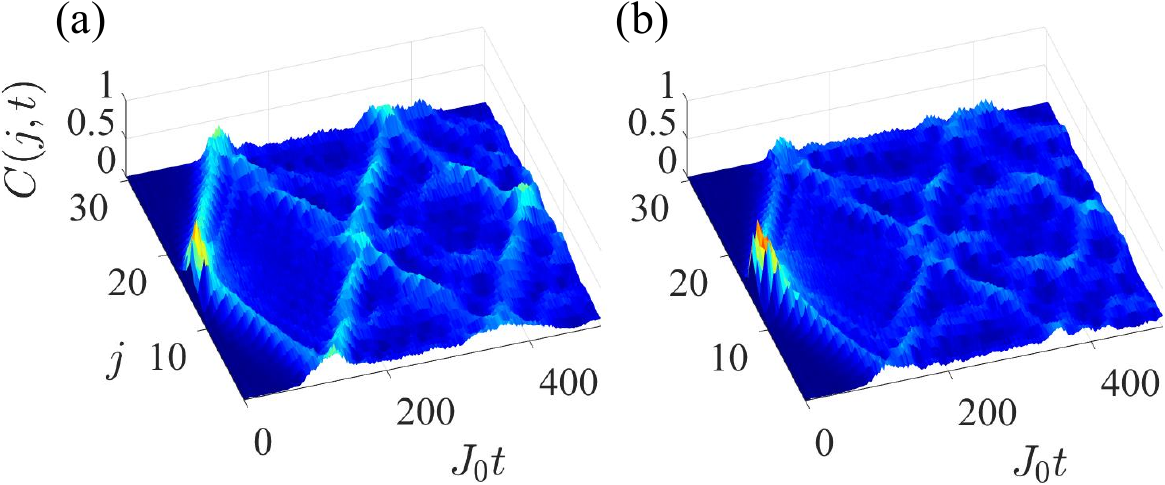}
    \caption{Time evolution of the normalized out-of-time-ordered commutator for maximally localized MWS in the (a) nearly linear band and (b) curved band. 
    Parameters are $J_0=1$, $\delta_U=0.01$, $U_0=20$, $d=3$, $g(j)\in\{-0.20, 0.48, -0.30\}$, and $M=30$.
    }
    \label{OTOC}
\end{figure}
In the periodic revival dynamics, the operator spreading remains concentrated along recurrent ballistic trajectories instead of spreading diffusively over the whole system.
To show this restriction, we calculate the out-of-time-ordered (OTO) commutator~\cite{larkin1969quasiclassical_SM,kitaev2015simple_SM,maldacena2016bound_SM,swingle2018unscrambling_SM,PRXQuantum.5.010201_SM,cg3f-rggs_SM} 
\begin{equation}
    C(j,t)=\langle \psi|[\hat{n}_j(t),\hat{n}_{15}]^{\dagger}[\hat{n}_j(t),\hat{n}_{15}]|\psi\rangle,
\end{equation}
a standard diagnostic of operator spreading and quantum scrambling, where $\hat{n}_j(t)=e^{i\hat{H}t}\hat{n}_je^{-i\hat{H}t}$ and $|\psi\rangle$ is the initial maximally localized MWS.
In the nearly linear band, the maximally localized MWS splits into wave packets with well-defined opposite group velocities, 
so the OTO commutator follows the same recurrent ballistic paths as the density dynamics [Fig.~\ref{OTOC}(a)], which reflects a limited quantum scrambling.
However, in the curved highest dimer-monomer band, different momentum components have different group velocities.
The OTO commutator signal therefore broadens over the system [Fig.~\ref{OTOC}(b)], and the ballistic trajectory becomes blurred.
Parameters are set as  $J_0=1$, $\delta_U=0.01$, $U_0=20$, $d=3$, $g(j)\in\{-0.20, 0.48, -0.30\}$, and $M=30$.
The values are normalized by $C(j,t)/{\rm max}(C(j,t))$.
This contrast indicates that the coherent revival dynamics not only preserves the initial memory, 
but also constrains operator spreading.

\subsection{Long-time dynamics in the superlattice and simple lattice}
\begin{figure}[!h]
    \centering
    \includegraphics[width=0.7\linewidth]{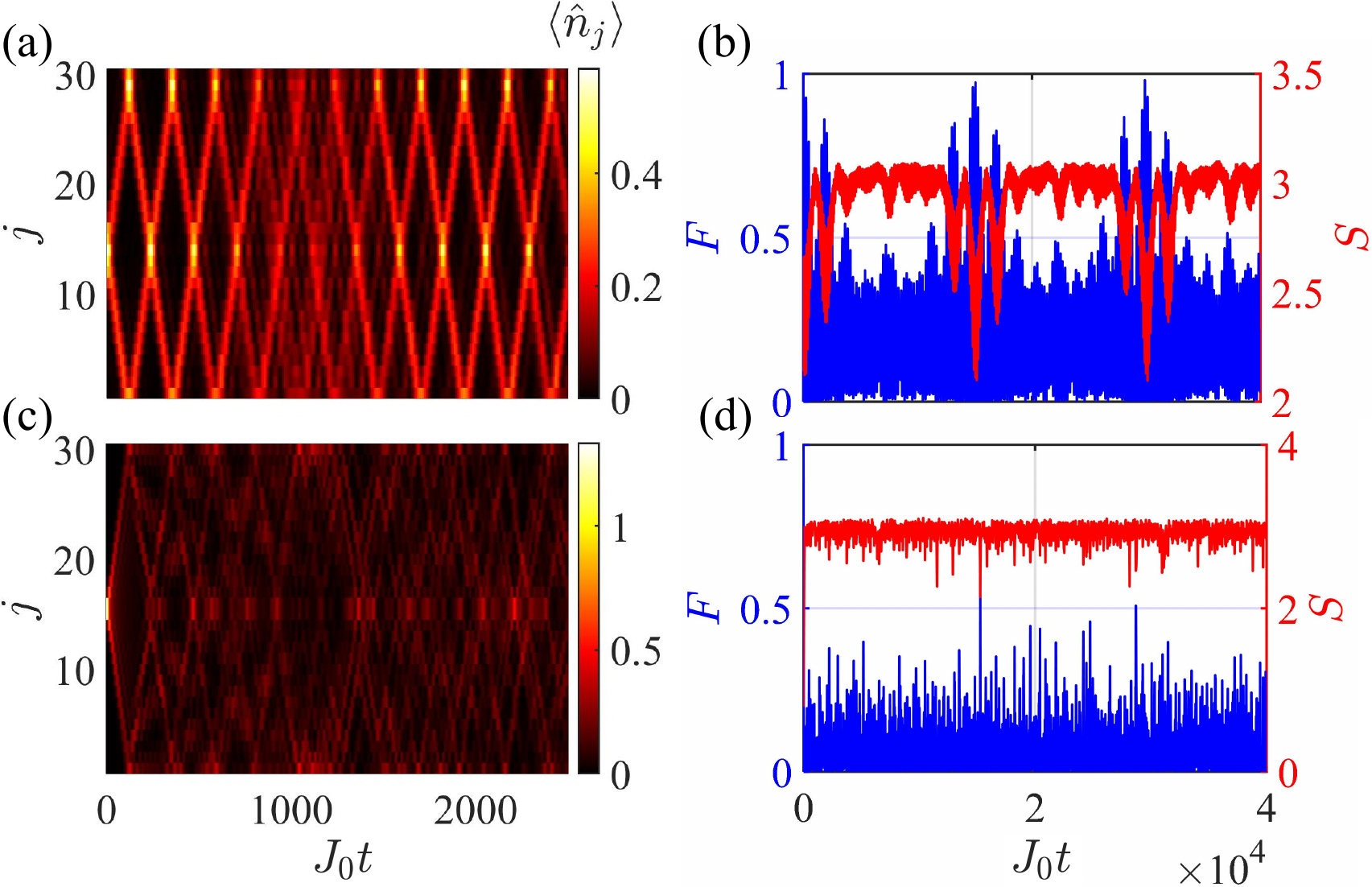}
    \caption{Long-time dynamics of dimer-monomer maximally localized MWS in the superlattice and simple lattice.
    Time-evolution of density distribution in (a) superlattice and (c) simple lattice. 
    Time-evolution of fidelity (blue) and entanglement entropy (red) in (b) superlattice and (d) simple lattice.
    Parameters are chosen as $J_0=1$, $U_0=20$, $d=3,$ $g(j)\in\{-0.20,0.48,-0.30\}$, $M=30$.
    $\delta_U=0.01$ for (a),(b) and $\delta_U=0$ for (c),(d).
    }
    \label{longtime}
\end{figure}
To complement Fig.~3 of the main text, we show here the long-time evolution of the maximally localized MWS in both the superlattice and the corresponding simple lattice.
For maximally localized MWS in the nearly linear band of superlattice,  
apart from the equal-spacing energy levels, the level spacing around the band edges exhibits slight deviations.
The dominant equal-spacing structure produces regular revivals, while the small deviations near the band edges introduce additional close frequencies, leading to long-time beat oscillations; see Figs.~\ref{longtime}(a),(b).
However, for maximally localized MWS in the curved highest dimer-monomer band of simple lattice, 
the coherence of dynamics is lost and the entanglement entropy rapidly increases and saturates; see Figs.~\ref{longtime}(c),(d).
Although apparent recurrences can still be observed in the finite-size simple lattice, 
they are irregular finite-size recurrences rather than coherent revivals protected by an equal-spacing structure.
When the system size increases, the apparent recurrences will be broken while the coherent revivals persist well; see Fig.~4 in the main text and the Section~\ref{finitesize}.

\subsection{Robustness of periodic revival dynamics against weak disorder}
\begin{figure}[!h]
    \centering
    \includegraphics[width=0.75\linewidth]{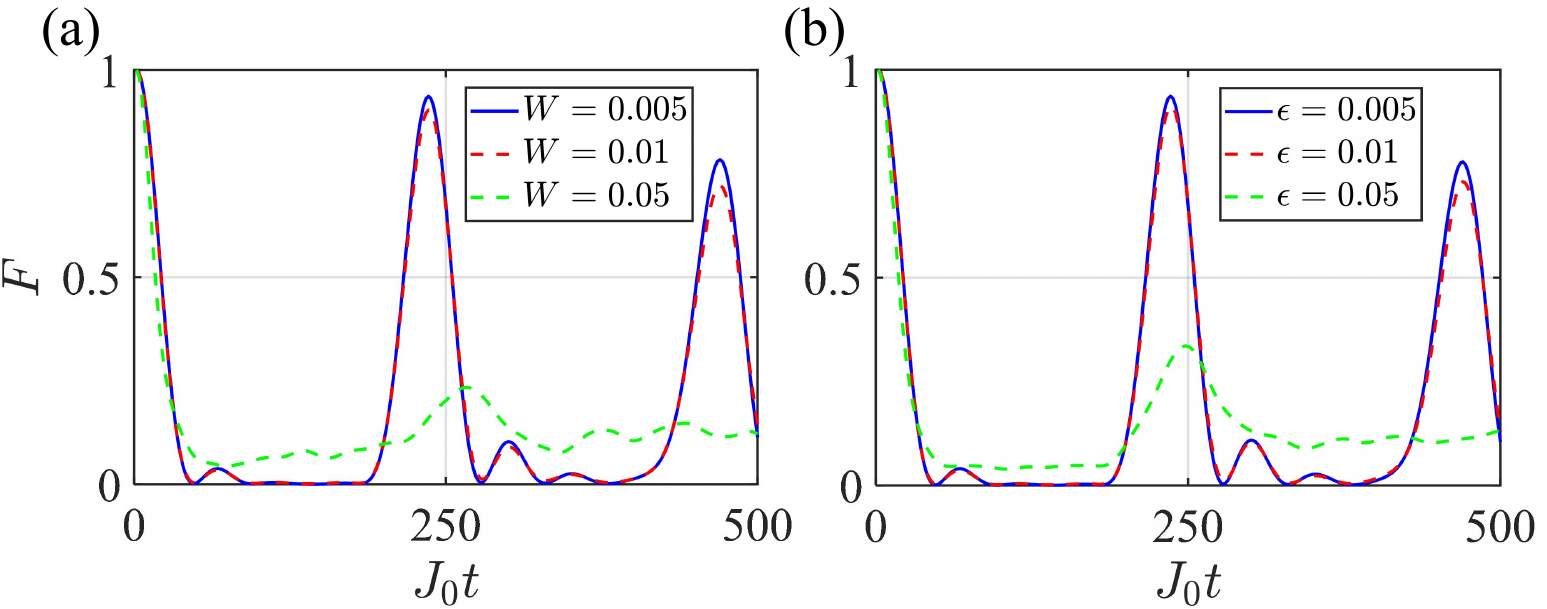}
    \caption{Disordered dynamics of the maximally localized MWS.
    Fidelity as a function of time with (a) onsite potential disorder strength $W=0.005$, $W=0.01$ and $W=0.05$, and (b) hopping disorder strength $\epsilon=0.005$, $\epsilon=0.01$ and $\epsilon=0.05$.
    Parameters are $J_0=1$, $\delta_U=0.01$, $U_0=20$, $d=3$, $g(j)\in\{-0.20, 0.48, -0.30\}$, $M=30$.
    The fidelity is averaged over $30$ disorder realizations. 
    }
    \label{robust}
\end{figure}
In this section, we present a robustness analysis for the periodic revival dynamics.
We first consider an external disorder in onsite energies, where the system is described by
\begin{equation}
    \hat{H}_{ d1}=\hat{H}+\sum_jW\mathcal{V}_j\hat{n}_j.
\end{equation}
Here, $\mathcal{V}_j$ are random values in $(-1,1)$, 
and $W$ is the disorder strength.
Considering a maximally localized MWS in the second highest dimer-monomer band as initial state,
Fig.~\ref{robust}(a) shows the fidelity between the evolved state and initial state with disorder strengths $W=0.005$, $0.01$ and $0.05$; see the blue solid, red dashed and green dashed lines, respectively.  
Besides, we also consider another type of disorder that is in the hopping strength.
The system is described by 
\begin{equation}
    \hat{H}_{d2}=\hat{H}+\epsilon \sum_j h_j(\hat{a}_{j+1}^{\dagger}\hat{a}_j+{\rm H.c.}),
\end{equation}
where the hopping strengths becomes $J_0+\epsilon h_j$, 
and $h_j$ are also random values in $(-1,1)$ which are independent from $\mathcal{V}_j$.
Fig.~\ref{robust}(b) shows the dynamics with $\epsilon=0.005$, $0.01$ and $0.05$.
Parameters are $J_0=1$, $\delta_U=0.01$, $U_0=20$, $d=3$, $g(j)\in\{-0.20, 0.48, -0.30\}$, $M=30$,
All processes are averaged over $30$ random realizations of $\mathcal{V}_j$ or $h_j$. 
Although disorder breaks exact cotranslation symmetry and can induce weak coupling between Wannier sectors, the revival remains visible when the induced level shifts and inter-sector couplings are sufficiently small.
These results demonstrate robustness against perturbations for the revival dynamics.

\subsection{Periodic revival dynamics from a Fock-state superposition}
\begin{figure}[!h]
    \centering
    \includegraphics[width=0.9\linewidth]{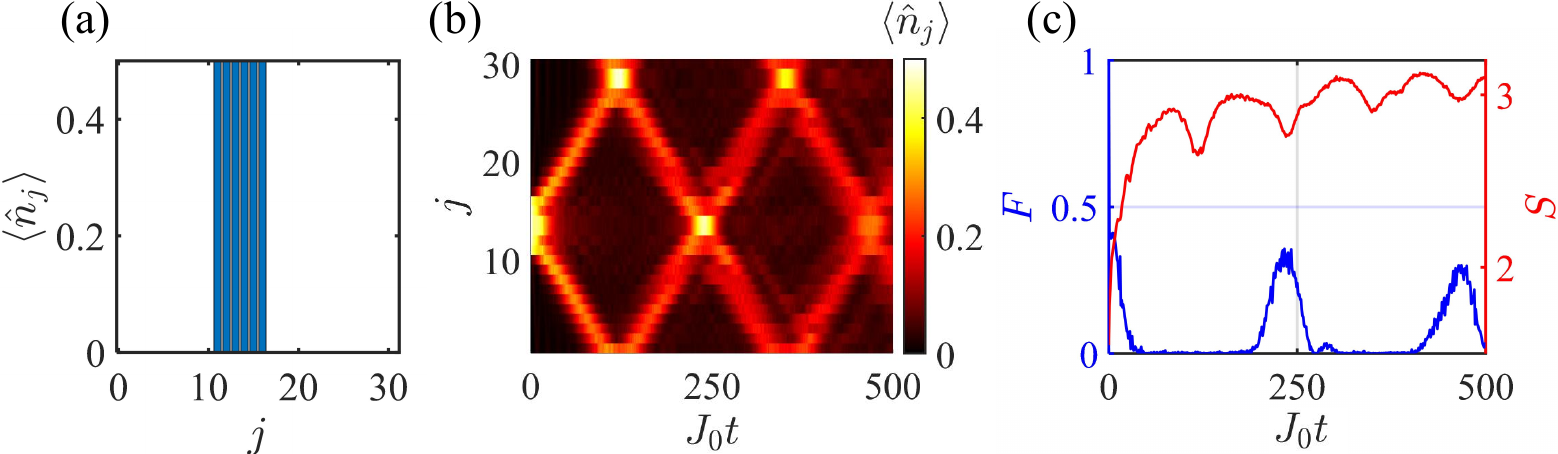}
    \caption{Periodic revival dynamics of the Fock states. 
    (a) The density distribution of the initial state.
    (b) Density distribution, (c) fidelity and entanglement entropy as a function of time. 
    Parameters are set as $J_0=1$, $\delta_U=0.01$, $U_0=20$, $d=3$, $g(j)\in\{-0.20, 0.48, -0.30\}$, and $M=30$.
    }
    \label{physical}
\end{figure}
In this section, we show the periodic revival dynamics a more experimentally accessible state than the exact maximally localized MWS.
Fig.~\ref{physical}(a) shows the density distribution of a selected initial state
\begin{equation}
    |\psi\rangle=\frac{1}{\sqrt{6}}(|2\rangle_{12}|1\rangle_{11}-|2\rangle_{11}|1\rangle_{12}+|2\rangle_{13}|1\rangle_{14}-|2\rangle_{14}|1\rangle_{13}+|2\rangle_{16}|1\rangle_{15}-|2\rangle_{15}|1\rangle_{16}),
\end{equation}
where $|2\rangle_i|1\rangle_j$ denote the Fock states $|0,...,n_i=2,...,n_j=1,...,0\rangle$.
In contrast to the generic dimer-monomer Fock states studied in Section~\ref{Fockdynamics}, 
the initial state considered here is a coherent superposition of a few Fock configurations chosen to have substantial overlap with the maximally localized MWS in the nearly linear band.
We calculate the time evolution of the density distribution, fidelity and entanglement entropy in the dynamics initiated from the initial state, as shown in Fig.~\ref{physical}(b) and (c) respectively.
The parameters are chosen as those in Fig.~3 of the main text.
While the initial state is not an exact maximally localized MWS, pronounced periodic revivals persist.
Therefore, exact preparation of an ideal maximally localized MWS is not required.
Pronounced revivals can be observed as long as the initial state has a substantial projection onto the coherent nearly linear Wannier sector.

\subsection{Finite-size recurrences versus coherent revivals}\label{finitesize}
\begin{figure}[!h]
    \centering
    \includegraphics[width=0.8\linewidth]{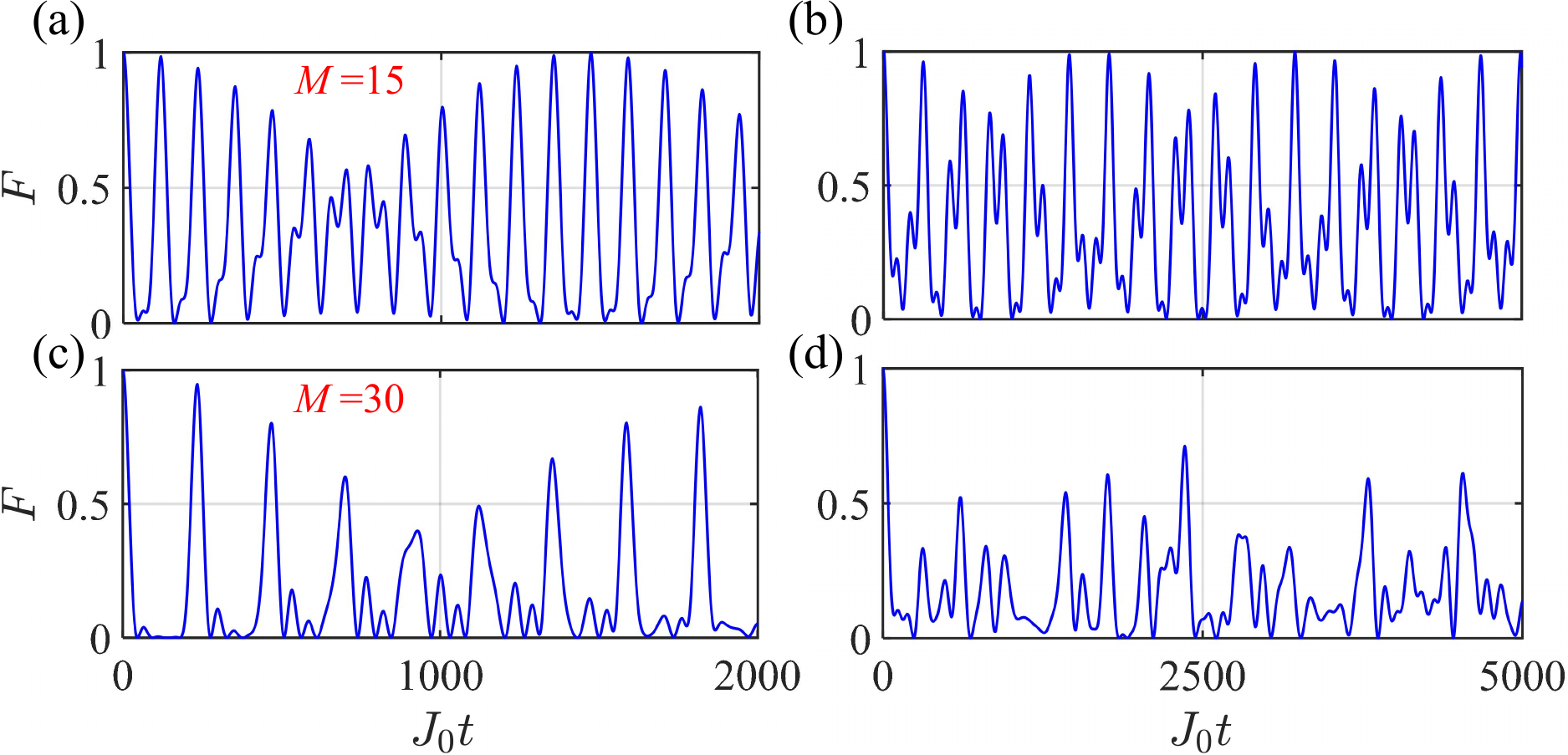}
    \caption{Fidelity as a function of time during the dynamics of maximally localized MWS in the nearly linear (left panel) and curved (right panel) dimer-monomer band.
    Parameters are chosen as $J_0=1$, $\delta_U=0.01$, $U_0=20$, $d=3$, $g(j)\in\{-0.20, 0.48, -0.30\}$.
    $M=15$ for (a),(b) and $M=30$ for (c),(d).
    }
    \label{systemsize}
\end{figure}
In this section, we distinguish coherent revivals induced by the nearly equal-spacing structure from ordinary finite-size recurrences.
Considering a maximally localized MWS in the nearly linear second-highest dimer-monomer and curved highest dimer-monomer band as the initial state, 
for the system size $M=15$, both of them show apparent recurrences because only a small number of energy levels participate in the dynamics; see Figs.~\ref{systemsize}(a),(b).
However, as the system size increases,
while the coherence maintains well in the nearly linear band [Fig.~\ref{systemsize}(c)], 
more incommensurate energy spacings participate in the dynamics of the curved band, causing rapid dephasing and suppressing the recurrence peaks.
see Fig.~\ref{systemsize}(d) and Fig.~4 of the main text.
Parameters are chosen as $J_0=1$, $\delta_U=0.01$, $U_0=20$, $d=3$, $g(j)\in\{-0.20, 0.48, -0.30\}$.
Therefore, by engineering nearly linear band, the level spacings remain coherence compared to those of a generic irregular band, allowing the revival dynamics to persist to larger system sizes.

\section{Periodic revival dynamics beyond the three-particle case}
\subsection{Periodic revival dynamics of $(N-1)$-bound-monomer state}
\label{more_particle}
\begin{figure}[!h]
    \centering
    \includegraphics[width=1\linewidth]{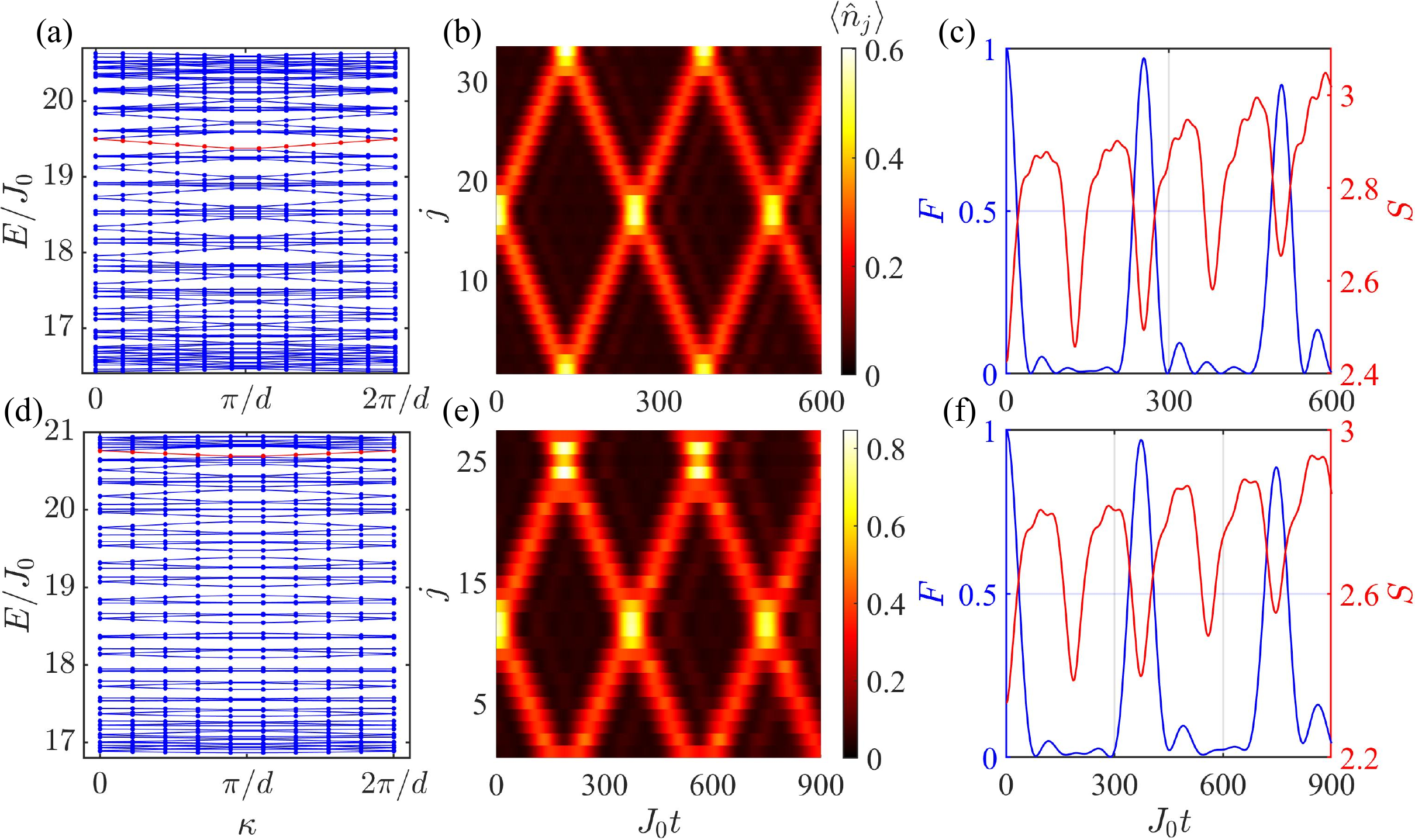}
    \caption{(a) Four-particle Bloch bands corresponding to three bound particles and an independent particle.
    (b) Density distribution, (c) fidelity (blue) and entanglement entropy (red) as a function of time when a maximally localized MWS in the linear band marked by red in (a) evolves. 
    (d) Five-particle Bloch bands correspond to four bound particles and an independent one.
    (e) Density distribution, (f) fidelity (blue) and entanglement entropy (red) as a function of time when a maximally localized MWS in the linear band marked by red in (d) evolves. 
    Parameters are $(J_0, V_0, U_0, \delta_J, \delta_V, \delta_U, d, M)=(1,0,6,0.01,0,0,3,33)$, $g(j)\in\{ 0.39,-0.37,0.90\}$ for (a),(b),(c) and $(J_0, V_0, U_0, \delta_J, \delta_V, \delta_U, d, M)=(1,0,3,0.01,0,0,3,27)$, $g(j)\in\{ -0.75, 0.63, 0.81\}$ for (d),(e),(f).
    }
    \label{moreparticle}
\end{figure}
The coherent-band mechanism is not restricted to the three-particle case.
As representative examples beyond three particles, we consider four- and five-particle systems.
Fig.~\ref{moreparticle}(a) shows the $(N-1)$-bound-monomer bands in the four-particle system, 
where three particles form bound states and one particle is independent.
Fig.~\ref{moreparticle}(d) shows the similar result in the five-particle system, 
where four particles form bound states and one particle is independent. 
There are a number of nearly linear bands embedded in the continuum of $(N-1)$-bound-monomer states under the moderate modulation of hopping strength.
Periodic revival dynamics occurs for the maximally localized MWSs in a nearly linear bands.
As examples, we consider maximally localized MWSs in the bands marked by red color in Fig.~\ref{moreparticle}(a) and Fig.~\ref{moreparticle}(d) as the initial states,
and the corresponding dynamics are shown in Figs.~\ref{moreparticle}(b),(c) and Figs.~\ref{moreparticle}(e),(f), respectively.
Both the four- and five-particle systems show pronounced coherent revival dynamics.
Parameters are 
$(J_0, V_0, U_0, \delta_J, \delta_V, \delta_U, d, M)=(1,0,6,0.01,0,0,3,33)$, $g(j)\in\{ 0.39,-0.37,0.90\}$ for Figs.~\ref{moreparticle}(a),(b),(c) 
and $(J_0, V_0, U_0, \delta_J, \delta_V, \delta_U, d, M)=(1,0,3,0.01,0,0,3,27)$, $g(j)\in\{  -0.75, 0.63, 0.81\}$ for Figs.~\ref{moreparticle}(d),(e),(f).
We note that small interaction strength is enough to induce the $(N-1)$-bound-monomer states for large number of particles,
because it is easier to be large enough energy $\frac{U_j}{2}N(N-1)$ for bound states with large number of particles.

\subsection{Periodic revival dynamics in other manifolds}
\begin{figure}[!h]
    \centering
    \includegraphics[width=1\linewidth]{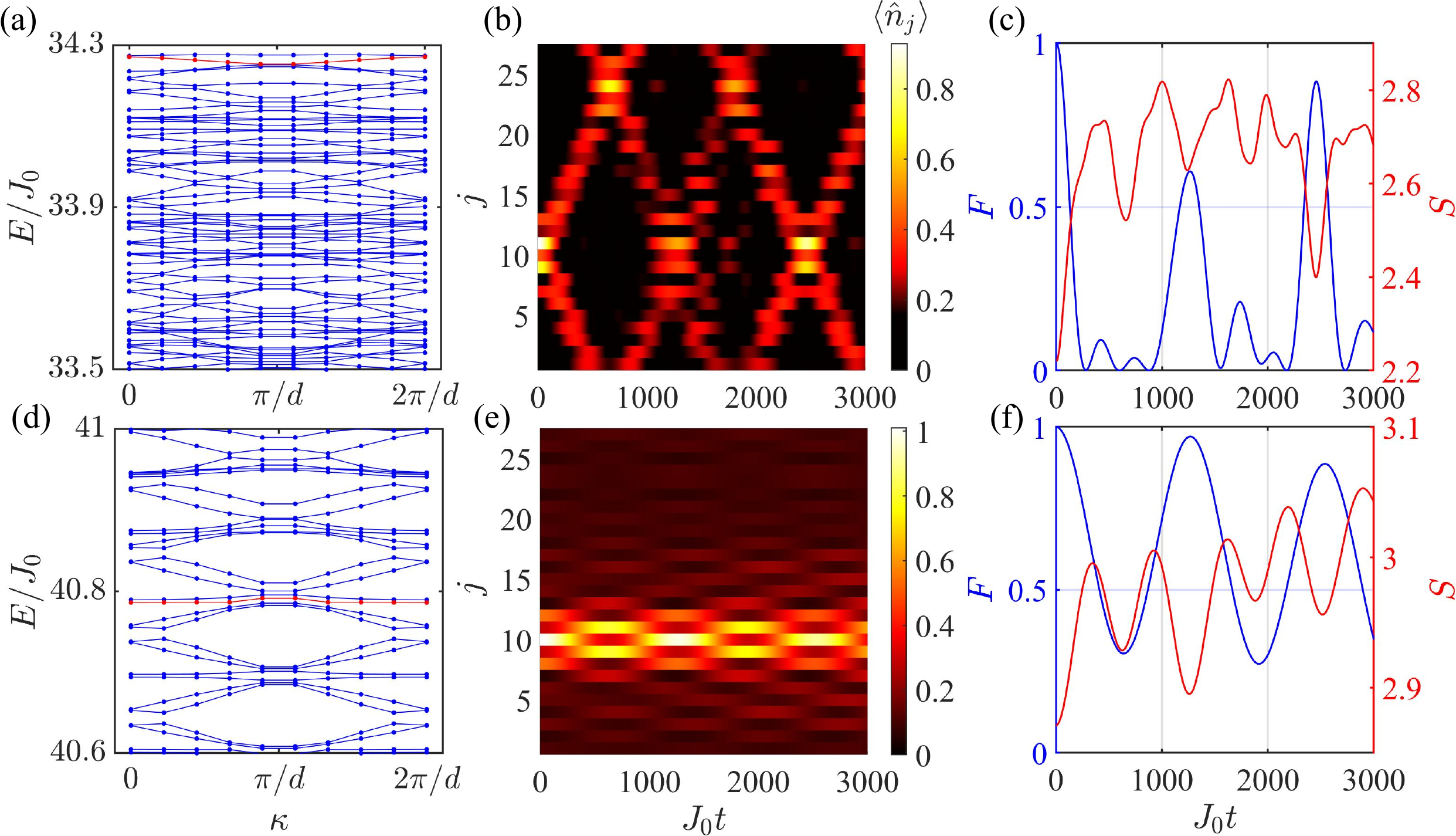}
    \caption{(a) Part of Bloch bands corresponding to three bound particles and two independent particle ($3+1+1$).
    (b) Density distribution, (c) fidelity (blue) and entanglement entropy (red) as a function of time when a maximally localized MWS in the band marked by red in (a) evolves. 
    (d) Part of Bloch bands corresponding to three bound particles and other two bound particles ($3+2$).
    (e) Density distribution, (f) fidelity (blue) and entanglement entropy (red) as a function of time when a maximally localized MWS in the band marked by red in (d) evolves. 
    Parameters are 
    $(J_0, V_0, U_0, \delta_J, \delta_V, \delta_U, d, M)=(1,0,10,0.01,0,0,3,27)$, $g(j)\in\{-0.75, 0.63, 0.81 \}$.
    }
    \label{othertype}
\end{figure}
The coherent-revival mechanism is not limited to the sector of the $(N-1)$ bound-monomer states.
Other hybrid sectors involving $(N-2)$-particle bound state plus $2$ monomer states, and $(N-2)$-particle bound state plus $2$- particle bound state, can also host coherent bands embedded in a thermal background and provide overall localized maximally localized MWS. 
For $N=5$, we consider 
(i) a three-particle bound state plus two independent particles,
and (ii) a three-particle bound state plus a two-particle bound state.
Fig.~\ref{othertype}(a) shows part of the bands of the former states.
We consider a nearly linear band marked by red, and periodic revival dynamics occurs for the maximally localized MWSs; see Figs.~\ref{othertype}(b),(c).
Fig.~\ref{othertype}(d) shows part of the bands of the latter states. 
As an example, we consider the band marked by red, there can also be periodic revivals for the maximally localized MWSs; see Figs.~\ref{othertype}(e),(f), respectively.
Parameters are 
$(J_0, V_0, U_0, \delta_J, \delta_V, \delta_U, d, M)=(1,0,10,0.01,0,0,3,27)$, $g(j)\in\{ -0.75,0.63, 0.81\}$.
These examples show that the mechanism is not limited to the specific configuration of $(N-1)$- bound-monomer states.

\end{CJK}

\begin{thebibliography}{77}
\bibitem{PhysRevA.43.2046} J. M. Deutsch, Quantum statistical mechanics in a closed system, \href{https://doi.org/10.1103/PhysRevA.43.2046}{Phys. Rev. A 43, 2046 (1991)}.
\bibitem{PhysRevE.50.888} M. Srednicki, Chaos and quantum thermalization, \href{https://doi.org/10.1103/PhysRevE.50.888}{Phys. Rev. E 50, 888 (1994)}.
\bibitem{srednicki1999approach} M. Srednicki, The approach to thermal equilibrium in quantized chaotic systems, \href{https://iopscience.iop.org/article/10.1088/0305-4470/32/7/007}{J. Phys. A:Math. Gen. 32, 1163 (1999)}.
\bibitem{rigol2008thermalization} M. Rigol, V. Dunjko, and M. Olshanii, Thermalization and its mechanism for generic isolated quantum systems, \href{https://www.nature.com/articles/nature06838}{Nature (London) 452, 854 (2008)}.
\bibitem{Deutsch_2018} J. M. Deutsch, Eigenstate thermalization hypothesis, \href{https://doi.org/10.1088/1361-6633/aac9f1}{Rep. Prog. Phys. 81, 082001 (2018)}.
\bibitem{kinoshita2006quantum} T. Kinoshita, T. Wenger, and D. S. Weiss, A quantum Newton's cradle, \href{https://www.nature.com/articles/nature04693}{Nature (London) 440, 900 (2006)}.
\bibitem{PhysRevLett.98.050405} M. Rigol, V. Dunjko, V. Yurovsky, and M. Olshanii, Relaxation in a Completely Integrable Many-Body Quantum System: An Ab Initio Study of the Dynamics of the Highly Excited States of 1D Lattice Hard-Core Bosons, \href{https://doi.org/10.1103/PhysRevLett.98.050405}{Phys. Rev. Lett. 98, 050405 (2007)}.
\bibitem{RevModPhys.83.863} A. Polkovnikov, K. Sengupta, A. Silva, and M. Vengalattore, Colloquium: Nonequilibrium dynamics of closed interacting quantum systems, \href{https://doi.org/10.1103/RevModPhys.83.863}{Rev. Mod. Phys. 83, 863 (2011)}.
\bibitem{PhysRevLett.95.206603} I. V. Gornyi, A. D. Mirlin, and D. G. Polyakov, Interacting Electrons in Disordered Wires: Anderson Localization and Low-T Transport, \href{https://doi.org/10.1103/PhysRevLett.95.206603}{Phys. Rev. Lett. 95, 206603 (2005)}.
\bibitem{BASKO20061126} D. Basko, I. Aleiner, and B. Altshuler, Metal--insulator transition in a weakly interacting many-electron system with localized single-particle states, \href{https://doi.org/10.1016/j.aop.2005.11.014}{Ann. Phys. 321, 1126 (2006)}.
\bibitem{ROS2015420} V. Ros, M. M\"uller, and A. Scardicchio, Integrals of motion in the many-body localized phase, \href{https://doi.org/10.1016/j.nuclphysb.2014.12.014}{Nucl. Phys. B 891, 420 (2015)}.
\bibitem{annurev:/content/journals/10.1146/annurev-conmatphys-031214-014726} R. Nandkishore and D. A. Huse, Many-Body Localization and Thermalization in Quantum Statistical Mechanics, \href{https://doi.org/10.1146/annurev-conmatphys-031214-014726}{Annu. Rev. Condens. Matter Phys. 6, 15 (2015)}.
\bibitem{annurev:/content/journals/10.1146/annurev-conmatphys-031214-014701} E. Altman and R. Vosk, Universal Dynamics and Renormalization in Many-Body-Localized Systems, \href{https://doi.org/10.1146/annurev-conmatphys-031214-014701}{Annu. Rev. Condens. Matter Phys. 6, 383 (2015)}.
\bibitem{RevModPhys.91.021001} D. A. Abanin, E. Altman, I. Bloch, and M. Serbyn, Colloquium: Many-body localization, thermalization, and entanglement, \href{https://doi.org/10.1103/RevModPhys.91.021001}{Rev. Mod. Phys. 91, 021001 (2019)}.
\bibitem{PhysRevLett.122.040606} M. Schulz, C. A. Hooley, R. Moessner, and F. Pollmann, Stark Many-Body Localization, \href{https://doi.org/10.1103/PhysRevLett.122.040606}{Phys. Rev. Lett. 122, 040606 (2019)}.
\bibitem{doi:10.1073/pnas.1819316116} E. van Nieuwenburg, Y. Baum, and G. Refael, From Bloch oscillations to many-body localization in clean interacting systems, \href{https://doi.org/10.1073/pnas.1819316116}{Proc. Natl. Acad. Sci. 116, 9269 (2019)}.
\bibitem{PhysRevB.102.054206} S. R. Taylor, M. Schulz, F. Pollmann, and R. Moessner, Experimental probes of Stark many-body localization, \href{https://doi.org/10.1103/PhysRevB.102.054206}{Phys. Rev. B 102, 054206 (2020)}.
\bibitem{PhysRevA.103.023323} L. Zhang, Y. Ke, W. Liu, and C. Lee, Mobility edge of Stark many-body localization, \href{https://doi.org/10.1103/PhysRevA.103.023323}{Phys. Rev. A 103, 023323 (2021)}.
\bibitem{PhysRevLett.127.240502} Q. Guo, C. Cheng, H. Li, S. Xu, P. Zhang, Z. Wang, C. Song, W. Liu, W. Ren, H. Dong, R. Mondaini, and H. Wang, Stark Many-Body Localization on a Superconducting Quantum Processor, \href{https://doi.org/10.1103/PhysRevLett.127.240502}{Phys. Rev. Lett. 127, 240502 (2021)}.
\bibitem{lhsz-dkmq} L. Zhang, Y. Ke, and C. Lee, Suppressing Floquet thermalization by driving transparency in tilted lattices, \href{https://doi.org/10.1103/lhsz-dkmq}{Phys. Rev. Res. 7, 033206 (2025)}.
\bibitem{bernien2017probing} H. Bernien, S. Schwartz, A. Keesling, H. Levine, A. Omran, H. Pichler, S. Choi, A. S. Zibrov, M. Endres, M. Greiner, et al., Probing many-body dynamics on a 51-atom quantum simulator, \href{https://www.nature.com/articles/nature24622}{Nature (London) 551, 579 (2017)}.
\bibitem{turner2018weak} C. J. Turner, A. A. Michailidis, D. A. Abanin, M. Serbyn, and Z. Papi\'c, Weak ergodicity breaking from quantum many-body scars, \href{https://www.nature.com/articles/s41567-018-0137-5}{Nat. Phys. 14, 745 (2018)}.
\bibitem{PhysRevB.98.155134} C. J. Turner, A. A. Michailidis, D. A. Abanin, M. Serbyn, and Z. Papi\'c, Quantum scarred eigenstates in a Rydberg atom chain: Entanglement, breakdown of thermalization, and stability to perturbations, \href{https://doi.org/10.1103/PhysRevB.98.155134}{Phys. Rev. B 98, 155134 (2018)}.
\bibitem{PhysRevLett.122.040603} W. W. Ho, S. Choi, H. Pichler, and M. D. Lukin, Periodic Orbits, Entanglement, and Quantum Many-Body Scars in Constrained Models: Matrix Product State Approach, \href{https://doi.org/10.1103/PhysRevLett.122.040603}{Phys. Rev. Lett. 122, 040603 (2019)}.
\bibitem{PhysRevX.10.011055} A. A. Michailidis, C. J. Turner, Z. Papi\'c, D. A. Abanin, and M. Serbyn, Slow Quantum Thermalization and Many-Body Revivals from Mixed Phase Space, \href{https://doi.org/10.1103/PhysRevX.10.011055}{Phys. Rev. X 10, 011055 (2020)}.
\bibitem{PhysRevX.11.021021} C. J. Turner, J.-Y. Desaules, K. Bull, and Z. Papi\'c, Correspondence Principle for Many-Body Scars in Ultracold Rydberg Atoms, \href{https://doi.org/10.1103/PhysRevX.11.021021}{Phys. Rev. X 11, 021021 (2021)}.
\bibitem{doi:10.1126/science.abg2530} D. Bluvstein, A. Omran, H. Levine, A. Keesling, G. Semeghini, S. Ebadi, T. T. Wang, A. A. Michailidis, N. Maskara, W. W. Ho, S. Choi, M. Serbyn, M. Greiner, V. Vuleti\'c, and M. D. Lukin, Controlling quantum manybody dynamics in driven Rydberg atom arrays, \href{https://doi.org/10.1126/science.abg2530}{Science 371, 1355 (2021)}.
\bibitem{serbyn2021quantum} M. Serbyn, D. A. Abanin, and Z. Papi\'c, Quantum manybody scars and weak breaking of ergodicity, \href{https://www.nature.com/articles/s41567-021-01230-2}{Nat. Phys. 17, 675 (2021)}.
\bibitem{jepsen2022long} P. N. Jepsen, Y. K. Lee, H. Lin, I. Dimitrova, Y. Margalit, W. W. Ho, and W. Ketterle, Long-lived phantom helix states in Heisenberg quantum magnets, \href{https://www.nature.com/articles/s41567-022-01651-7}{Nat. Phys. 18, 899 (2022)}.
\bibitem{zhang2023many} P. Zhang, H. Dong, Y. Gao, L. Zhao, J. Hao, J.-Y. Desaules, Q. Guo, J. Chen, J. Deng, B. Liu, et al., Manybody Hilbert space scarring on a superconducting processor, \href{https://www.nature.com/articles/s41567-022-01784-9}{Nat. Phys. 19, 120 (2023)}.
\bibitem{PhysRevResearch.5.023010} G.-X. Su, H. Sun, A. Hudomal, J.-Y. Desaules, Z.-Y. Zhou, B. Yang, J. C. Halimeh, Z.-S. Yuan, Z. Papi\'c, and J.-W. Pan, Observation of many-body scarring in a BoseHubbard quantum simulator, \href{https://doi.org/10.1103/PhysRevResearch.5.023010}{Phys. Rev. Res. 5, 023010 (2023)}.
\bibitem{annurev:/content/journals/10.1146/annurev-conmatphys-031620-101617} A. Chandran, T. Iadecola, V. Khemani, and R. Moessner, Quantum Many-Body Scars: A Quasiparticle Perspective, \href{https://doi.org/10.1146/annurev-conmatphys-031620-101617}{Annu. Rev. Condens. Matter Phys. 14, 443 (2023)}.
\bibitem{PhysRevLett.132.150401} H.-R. Wang, D. Yuan, S.-Y. Zhang, Z. Wang, D.-L. Deng, and L.-M. Duan, Embedding Quantum Many-Body Scars into Decoherence-Free Subspaces, \href{https://doi.org/10.1103/PhysRevLett.132.150401}{Phys. Rev. Lett. 132, 150401 (2024)}.
\bibitem{pizzi2025genuine} A. Pizzi, L.-H. Kwan, B. Evrard, C. B. Dag, and J. Knolle, Genuine quantum scars in many-body spin systems, \href{https://www.nature.com/articles/s41467-025-61765-3}{Nat. Commun. 16, 6722 (2025)}.
\bibitem{HanPu2025044600} H. Pu, Unusual quantum many-body scars, \href{https://doi.org/10.15302/frontphys.2025.044600}{Front. Phys. 20, 044600 (2025)}.
\bibitem{PhysRevX.9.021003} S. Pai, M. Pretko, and R. M. Nandkishore, Localization in Fractonic Random Circuits, \href{https://doi.org/10.1103/PhysRevX.9.021003}{Phys. Rev. X 9, 021003 (2019)}.
\bibitem{hudomal2020quantum} A. Hudomal, I. Vasi\'c, N. Regnault, and Z. Papi\'c, Quantum scars of bosons with correlated hopping, \href{https://www.nature.com/articles/s42005-020-0364-9}{Commun. Phys. 3, 99 (2020)}.
\bibitem{PhysRevX.10.011047} P. Sala, T. Rakovszky, R. Verresen, M. Knap, and F. Pollmann, Ergodicity Breaking Arising from Hilbert Space Fragmentation in Dipole-Conserving Hamiltonians, \href{https://doi.org/10.1103/PhysRevX.10.011047}{Phys. Rev. X 10, 011047 (2020)}.
\bibitem{PhysRevB.101.174204} V. Khemani, M. Hermele, and R. Nandkishore, Localization from Hilbert space shattering: From theory to physical realizations, \href{https://doi.org/10.1103/PhysRevB.101.174204}{Phys. Rev. B 101, 174204 (2020)}.
\bibitem{doi:10.1142/9789811231711_0009} S. Moudgalya, A. Prem, R. Nandkishore, N. Regnault, and B. A. Bernevig, Thermalization and Its Absence within Krylov Subspaces of a Constrained Hamiltonian, \href{https://doi.org/10.1142/9789811231711_0009}{in Memorial Volume for Shoucheng Zhang (World Scientific, 2022) Chap. Chapter 7, pp. 147--209}.
\bibitem{PhysRevX.12.011050} S. Moudgalya and O. I. Motrunich, Hilbert Space Fragmentation and Commutant Algebras, \href{https://doi.org/10.1103/PhysRevX.12.011050}{Phys. Rev. X 12, 011050 (2022)}.
\bibitem{moudgalya2022quantum} S. Moudgalya, B. A. Bernevig, and N. Regnault, Quantum many-body scars and Hilbert space fragmentation: a review of exact results, \href{https://iopscience.iop.org/article/10.1088/1361-6633/ac73a0}{Rep. Prog. Phys. 85, 086501 (2022)}.
\bibitem{PhysRevB.109.184313} L. Zhang, Y. Ke, L. Lin, and C. Lee, Floquet engineering of Hilbert space fragmentation in Stark lattices, \href{https://doi.org/10.1103/PhysRevB.109.184313}{Phys. Rev. B 109, 184313 (2024)}.
\bibitem{Li} L. Zhang, Y. Ke, and C. Lee, Zero-energy quantum many-body scar under emergent chiral symmetry and pseudo Hilbert space fragmentation, \href{https://doi.org/10.15302/frontphys.2025.044201}{Front. Phys. 20, 044201 (2025)}.
\bibitem{PhysRevX.15.011035} L. Zhao, P. R. Datla, W. Tian, M. M. Aliyu, and H. Loh, Observation of Quantum Thermalization Restricted to Hilbert Space Fragments and Z2k Scars, \href{https://doi.org/10.1103/PhysRevX.15.011035}{Phys. Rev. X 15, 011035 (2025)}.
\bibitem{yang2025constructing} F. Yang, M. Magoni, and H. Pichler, Constructing Quantum Many-Body Scars from Hilbert Space Fragmentation, \href{https://export.arxiv.org/abs/2506.10806}{arXiv:2506.10806 (2025)}.
\bibitem{7j6x-74f1} S. Aditya, Diagnostics of hilbert space fragmentation, freezing transition, and its effects in the family of quantum east models involving varying range of constraints, \href{https://doi.org/10.1103/7j6x-74f1}{Phys. Rev. B 112, 195413 (2025)}.
\bibitem{PhysRevResearch.7.023099} H. Katsura, C. Matsui, C. Paletta, and B. Pozsgay, Weak ergodicity breaking with isolated integrable sectors, \href{https://doi.org/10.1103/PhysRevResearch.7.023099}{Phys. Rev. Res. 7, 023099 (2025)}.
\bibitem{PhysRevB.106.035123} A. Russomanno, M. Fava, and R. Fazio, Weak ergodicity breaking in josephson-junction arrays, \href{https://doi.org/10.1103/PhysRevB.106.035123}{Phys. Rev. B 106, 035123 (2022)}.
\bibitem{PhysRevA.95.063630} Y. Ke, X. Qin, Y. S. Kivshar, and C. Lee, Multiparticle Wannier states and Thouless pumping of interacting bosons, \href{https://doi.org/10.1103/PhysRevA.95.063630}{Phys. Rev. A 95, 063630 (2017)}.
\bibitem{PhysRevB.96.195134} X. Qin, F. Mei, Y. Ke, L. Zhang, and C. Lee, Topological magnon bound states in periodically modulated Heisenberg XXZ chains, \href{https://doi.org/10.1103/PhysRevB.96.195134}{Phys. Rev. B 96, 195134 (2017)}.
\bibitem{qin2018topological} X. Qin, F. Mei, Y. Ke, L. Zhang, and C. Lee, Topological invariant and cotranslational symmetry in strongly interacting multi-magnon systems, \href{https://iopscience.iop.org/article/10.1088/1367-2630/aa9556}{New J. Phys. 20, 013003 (2018)}.
\bibitem{PhysRevA.101.023620} L. Lin, Y. Ke, and C. Lee, Interaction-induced topological bound states and Thouless pumping in a onedimensional optical lattice, \href{https://doi.org/10.1103/PhysRevA.101.023620}{Phys. Rev. A 101, 023620 (2020)}.
\bibitem{PhysRevResearch.5.013020} W. Liu, S. Hu, L. Zhang, Y. Ke, and C. Lee, Correlated topological pumping of interacting bosons assisted by Bloch oscillations, \href{https://doi.org/10.1103/PhysRevResearch.5.013020}{Phys. Rev. Res. 5, 013020 (2023)}.
\bibitem{huang2024topological} B. Huang, Y. Ke, W. Liu, and C. Lee, Topological pumping induced by spatiotemporal modulation of interaction, \href{https://iopscience.iop.org/article/10.1088/1402-4896/ad491e}{Phys. Scr. 99, 065997 (2024)}.
\bibitem{Supplementary} See Supplemental Material for details of (S1) Bandresolved Wannier-sector structure; (S2) Engineering nearly linear multiparticle sub-bands; (S3) Thermalizing background and eigenstate diagnostics; (S4) Periodic revival dynamics and stability; (S5) Periodic revival dynamics beyond the three-particle case; which includes Refs.[47, 73-78].
\bibitem{walter2023quantization} A.-S. Walter, Z. Zhu, M. G\"achter, J. Minguzzi, S. Roschinski, K. Sandholzer, K. Viebahn, and T. Esslinger, Quantization and its breakdown in a Hubbard--Thouless pump, \href{https://www.nature.com/articles/s41567-023-02145-w}{Nat. Phys. 19, 1471 (2023)}.
\bibitem{ke2023topological} Y. Ke and C. Lee, Topological quantum tango, \href{https://www.nature.com/articles/s41567-023-02169-2}{Nat. Phys. 19, 1387 (2023)}.
\bibitem{ZiyuTao2025033202} Z. Tao, W. Huang, J. Niu, L. Zhang, Y. Ke, X. Gu, L. Lin, J. Qiu, X. Sun, X. Yang, et al., Emulating Thouless pumping in the interacting Rice-Mele model using superconducting qutrits, \href{https://doi.org/10.15302/frontphys.2025.033202}{Front. Phys. 20, 033202 (2025)}.
\bibitem{winkler2006repulsively} K. Winkler, G. Thalhammer, F. Lang, R. Grimm, J. Hecker Denschlag, A. Daley, A. Kantian, H. B\"uchler, and P. Zoller, Repulsively bound atom pairs in an optical lattice, \href{https://www.nature.com/articles/nature04918}{Nature (London) 441, 853 (2006)}.
\bibitem{valiente2008two} M. Valiente and D. Petrosyan, Two-particle states in the Hubbard model, \href{https://iopscience.iop.org/article/10.1088/0953-4075/41/16/161002}{J. Phys. B:At., Mol. Opt. Phys. 41, 161002 (2008)}.
\bibitem{fukuhara2013microscopic} T. Fukuhara, P. Schau\ss{}, M. Endres, S. Hild, M. Cheneau, I. Bloch, and C. Gross, Microscopic observation of magnon bound states and their dynamics, \href{https://www.nature.com/articles/nature12541}{Nature (London) 502, 76 (2013)}.
\bibitem{zhang2023stable} N. Zhang, Y. Ke, L. Lin, L. Zhang, and C. Lee, Stable interaction-induced Anderson-like localization embedded in standing waves, \href{https://iopscience.iop.org/article/10.1088/1367-2630/acca9c}{New J. Phys. 25, 043021 (2023)}.
\bibitem{PhysRevLett.133.140202} B. Huang, Y. Ke, H. Zhong, Y. S. Kivshar, and C. Lee, Interaction-Induced Multiparticle Bound States in the Continuum, \href{https://doi.org/10.1103/PhysRevLett.133.140202}{Phys. Rev. Lett. 133, 140202 (2024)}.
\bibitem{PhysRevLett.133.193001} Y. Liu and S. Chen, Fate of Two-Particle Bound States in the Continuum in Non-Hermitian Systems, 
\href{https://doi.org/10.1103/PhysRevLett.133.193001}
{Phys. Rev. Lett. 133, 193001 (2024)}.
\bibitem{Kolovsky2004QuantumCI} A. R. Kolovsky and A. Buchleitner, Quantum chaos in the bose-hubbard model, \href{https://api.semanticscholar.org/CorpusID:119348652}{EPL 68, 632 (2004)}.
\bibitem{PhysRevB.75.155111} V. Oganesyan and D. A. Huse, Localization of interacting fermions at high temperature, \href{https://doi.org/10.1103/PhysRevB.75.155111}{Phys. Rev. B 75, 155111 (2007)}.
\bibitem{PhysRevLett.110.084101} Y. Y. Atas, E. Bogomolny, O. Giraud, and G. Roux, Distribution of the ratio of consecutive level spacings in random matrix ensembles, \href{https://doi.org/10.1103/PhysRevLett.110.084101}{Phys. Rev. Lett. 110, 084101 (2013)}.
\bibitem{Kollath_2010} C. Kollath, G. Roux, G. Biroli, and A. M. L\"auchli, Statistical properties of the spectrum of the extended bose--hubbard model, \href{https://doi.org/10.1088/1742-5468/2010/08/P08011}{Journal of Statistical Mechanics: Theory and Experiment 2010, P08011 (2010)}.
\bibitem{Page1993} D. N. Page, Average entropy of a subsystem, \href{https://doi.org/10.1103/PhysRevLett.71.1291}{Phys. Rev. Lett. 71, 1291 (1993)}.
\bibitem{poshakinskiy2021quantum} A. V. Poshakinskiy, J. Zhong, Y. Ke, N. A. Olekhno, C. Lee, Y. S. Kivshar, and A. N. Poddubny, Quantum Hall phases emerging from atom--photon interactions, \href{https://www.nature.com/articles/s41534-021-00372-8}{npj Quantum Inf. 7, 34 (2021)}.
\bibitem{PhysRevLett.130.120401} S. Ghosh, I. Paul, and K. Sengupta, Prethermal fragmentation in a periodically driven fermionic chain, \href{https://doi.org/10.1103/PhysRevLett.130.120401}{Phys. Rev. Lett. 130, 120401 (2023)}.
\bibitem{larkin1969quasiclassical} A. I. Larkin and Y. N. Ovchinnikov, Quasiclassical method in the theory of superconductivity, \href{https://jetp.ras.ru/cgi-bin/dn/e_028_06_1200.pdf}{Sov Phys JETP 28, 1200 (1969)}.
\bibitem{kitaev2015simple} A. Kitaev, A simple model of quantum holography, \href{https://online.kitp.ucsb.edu/online/entangled15/kitaev/}{in Talks at KITP (2015)}.
\bibitem{maldacena2016bound} J. Maldacena, S. H. Shenker, and D. Stanford, A bound on chaos, \href{https://link.springer.com/article/10.1007/JHEP08(2016)106#citeas}{Journal of High Energy Physics 2016, 106 (2016)}.
\bibitem{swingle2018unscrambling} B. Swingle, Unscrambling the physics of out-of-timeorder correlators, \href{https://www.nature.com/articles/s41567-018-0295-5}{Nature Physics 14, 988 (2018)}.
\bibitem{PRXQuantum.5.010201} S. Xu and B. Swingle, Scrambling dynamics and out-oftime-ordered correlators in quantum many-body systems, \href{https://doi.org/10.1103/PRXQuantum.5.010201}{PRX Quantum 5, 010201 (2024)}.
\bibitem{cg3f-rggs} Y.-C. Li, T.-G. Zhou, S. Zhang, Z. Wu, L. Zhao, H. Yin, X. An, H. Zhai, P. Zhang, X. Peng, and J. Du, Errorresilient reversal of quantum chaotic dynamics enabled by scramblons, \href{https://doi.org/10.1103/cg3f-rggs}{Phys. Rev. Lett. 136, 060403 (2026)}.
\end{thebibliography}

\begin{thebibliography}{7}
\bibitem{7j6x-74f1_SM} S. Aditya, Diagnostics of hilbert space fragmentation, freezing transition, and its effects in the family of quantum east models involving varying range of constraints, \href{https://doi.org/10.1103/7j6x-74f1}{Phys. Rev. B 112, 195413 (2025)}.
\bibitem{larkin1969quasiclassical_SM} A. I. Larkin and Y. N. Ovchinnikov, Quasiclassical method in the theory of superconductivity, \href{https://jetp.ras.ru/cgi-bin/dn/e_028_06_1200.pdf}{Sov Phys JETP 28, 1200 (1969)}.
\bibitem{kitaev2015simple_SM} A. Kitaev, A simple model of quantum holography, \href{https://online.kitp.ucsb.edu/online/entangled15/kitaev/}{in Talks at KITP (2015)}.
\bibitem{maldacena2016bound_SM} J. Maldacena, S. H. Shenker, and D. Stanford, A bound on chaos, \href{https://link.springer.com/article/10.1007/JHEP08(2016)106#citeas}{Journal of High Energy Physics 2016, 106 (2016)}.
\bibitem{swingle2018unscrambling_SM} B. Swingle, Unscrambling the physics of out-of-time-order correlators, \href{https://www.nature.com/articles/s41567-018-0295-5}{Nature Physics 14, 988 (2018)}.
\bibitem{PRXQuantum.5.010201_SM} S. Xu and B. Swingle, Scrambling dynamics and out-of-time-ordered correlators in quantum many-body systems, \href{https://doi.org/10.1103/PRXQuantum.5.010201}{PRX Quantum 5, 010201 (2024)}.
\bibitem{cg3f-rggs_SM} Y.-C. Li, T.-G. Zhou, S. Zhang, Z. Wu, L. Zhao, H. Yin, X. An, H. Zhai, P. Zhang, X. Peng, and J. Du, Error-resilient reversal of quantum chaotic dynamics enabled by scramblons, \href{https://doi.org/10.1103/cg3f-rggs}{Phys. Rev. Lett. 136, 060403 (2026)}.
\end{thebibliography}
\end{document}